\documentclass[usenatbib]{mnras}

\usepackage{newtxtext,newtxmath}
\usepackage[T1]{fontenc}

\DeclareRobustCommand{\VAN}[3]{#2}
\let\VANthebibliography\thebibliography
\def\thebibliography{\DeclareRobustCommand{\VAN}[3]{##3}\VANthebibliography}

\usepackage{graphicx}	% Including figure files
\usepackage{amsmath}	% Advanced maths commands
\usepackage{float}
\usepackage[normalem]{ulem}
\usepackage{xspace,color}

\usepackage{soul,ulem}

\title[Chemical Abundance of LINER galaxies]{Chemical abundance of LINER galaxies - Metallicity calibrations based on SDSS-IV MaNGA}

\author[C.B. Oliveira Jr. et al.]{
C. B. Oliveira Jr.,$^{1}$\thanks{E-mail: celsojr@univap.br (CBO)}
A.C. Krabbe,$^{1}$\thanks{E-mail: angela.krabbe@gmail.com (ACK)}
J. A. Hernandez-Jimenez,$^{1}$
O. L. Dors Jr.,$^{1}$
\newauthor{I. A. Zinchenko,$^{2,3}$
G.~F. H\"agele,$^{4,5}$
 M.~V. Cardaci,$^{4,5}$
  A. F. Monteiro,$^{1,6}$}\\
$^{1}$ Universidade do Vale do Para\'{\i}ba, Av. Shishima Hifumi, 2911, Zip Code 12244-000, S\~ao Jos\'e dos Campos, SP, Brazil\\
$^2$ Faculty of Physics, Ludwig-Maximilians-Universit\"{a}t, Scheinerstr. 1, 81679 Munich, Germany \\
$^3$ Main Astronomical Observatory, National Academy of Sciences of Ukraine, 27 Akad. Zabolotnoho St 03680 Kyiv, Ukraine \\
$^{4}$ Instituto de Astrof\'isica de La Plata (CONICET La Plata--UNLP), Argentina. \\
$^{5}$ Facultad de Ciencias Astron\'omicas y Geof\'{\i}sicas, Universidad Nacional de La Plata, Paseo del Bosque s/n, 1900 La Plata, Argentina\\
$^{6}$ Instituto Federal do Maranh{\~a}o, Av. Newton Bello, s/n, Zip Code 65906-335,  Imperatriz, MA, Brazil\\
}

\date{Accepted XXX. Received YYY; in original form ZZZ}

\pubyear{2021}

\begin{document}
\label{firstpage}
\pagerange{\pageref{firstpage}--\pageref{lastpage}}
\maketitle
% Abstract of the paper
\begin{abstract}{The ionizing source of Low Ionization Nuclear Emission Regions (LINERs)  is uncertain. Because of this,  
an empirical relation to determine the chemical abundances of these objects has not been proposed. 
In this work, for the first time, we derived  two semi-empirical calibrations based on photoionization models to estimate the oxygen abundance of LINERS  
as a  function of the $N2$ and $O3N2$ emission-line intensity ratios. These relations were calibrated using oxygen abundance estimations obtained by comparing   the observational emission-line ratios of 43 LINER galaxies (taken from the MaNGA survey) and grids of photoionization models built with the {\sc Cloudy} code assuming post-Asymptotic Giant Branch (post-AGB) stars with different temperatures. We found that the oxygen abundance of LINERs in our sample is in the  $\rm 8.48 \: \la \: 12+log(O/H) \: \la  8.84$ range, with a mean value of $\rm 12+\log(O/H)=8.65$. 
We recommend the use of the $N2$ index to estimate the oxygen abundances of LINERs, since
the calibration with this index presented a much smaller dispersion than the $O3N2$ index.
In addition, the estimated metallicities are in good agreement with those derived by extrapolating the disk oxygen abundance gradients to the centre of the galaxies showing that the assumptions of the models are suitable for LINERs. We also obtained a calibration  between the logarithm of the ionization parameter and the  [\ion{O}{iii}]/[\ion{O}{ii}] emission-line ratio. %From this calibration, we obtained the ionization parameter for the studied LINERs and compared them with a sample of Seyfert galaxies. As expected, the LINERs have systematically lower values of the ionization parameter.
%The uncertain nature of the ionizing source of the gas phase of Low Ionization Nuclear Emission Regions (LINERs)
%that there is not still an empirical  
%to the determination of chemical abundances of these objects, until now there not empirical calibration

%un there is not a empirical calibration for

%we proposed, for the first time, two semi-empirical calibrations based on photoionization models to estimate the oxygen abundance of this class of objects, as a  function of the $N2$ and $O3N2$ emission-line intensity ratios 

%The uncertain nature of the ionizing source of the gas phase of Low Ionization Nuclear Emission Regions (LINERs)
%makes that there is not still an empirical relation to determine the  chemical abundances of these objects

%In this paper, we present two relations between the oxygen abundance and the  $N2$ and $O3N2$ indexes.

}

\end{abstract}
% Select between one and six entries from the list of approved keywords.
% Don't make up new ones.

%%%%%%%%%%%%%%%%%%%%%%%%%%%%%%%%%%%%%%%%%%%%%%%%%%

%%%%%%%%%%%%%%%%% BODY OF PAPER %%%%%%%%%%%%%%%%%%

\section{Introduction}

Chemical abundance  is a fundamental parameter to understanding the formation and evolution of galaxies. The metallicity ($Z$), i.e., the content of metals relative to hydrogen, is usually traced and parametrized  by the oxygen abundance relative to the hydrogen  (O/H)  of the gas phase,  since  oxygen is one of the most abundant elements produced after  primordial nucleosynthesis. The metallicity of the gas phase of Star Forming  regions  (SFs) and Active Galactic Nuclei (AGNs) can mainly be estimated using two methods. The first method, called $T_{\rm e}$-method, is based on direct determinations of electron temperatures (for a review see \citealt{2017PASP..129h2001P} and  \citealt{2017PASP..129d3001P}) and requires measurements of auroral emission-lines such as [\ion{O}{iii}]$\lambda$ 4363 and [\ion{N}{ii}]$\lambda$ 5755, which are generally weak or not measurable in objects with high metallicity or low excitation \citep{1998AJ....116.2805V, 2007MNRAS.382..251D,2008A&A...482...59D}. This technique is used to derive the gas-phase metallicity in SFs and  is widely accepted as producing the most reliable oxygen abundance estimates \citep{2003A&A...399.1003P,2006MNRAS.372..293H, 2008MNRAS.383..209H,2017MNRAS.467.3759T}. 
The second method is used when it is not possible to measure the auroral lines and the $T_{\rm e}$-method cannot be applied. 
Known  as the strong-line method or indirect method, as suggest by \cite{10.1093/mnras/189.1.95}, who followed the original idea of \cite{1976ApJ...209..748J}, this method is based on calibrations between the oxygen abundance (or metallicity) and strong emission-lines easily measured in SF region spectra (for a review see \citealt{2010A&A...517A..85L}, \citealt{2019A&ARv..27....3M} and  \citealt{2019ARA&A..57..511K}). 
 %In particular, studies based on strong-line methods have suggested that Seyfert 2 nuclei in the local universe ($z \: < \: 0.4$) present similar metallicities (or abundances) than those in metal rich \ion{H}{ii} regions, i.e., no extraordinary enrichment has been observed in AGNs, having these objects solar
%or slightly over-solar metallicities. This result agrees with predictions of chemical evolution
%models for spiral and elliptical galaxies (e.g. \citealt{molla}).

For oxygen abundances in AGNs,
 narrow line regions (NLRs) of Seyfert 2 are  by far  the most studied,  using both the $T_{\rm e}$-method (\citealt{1992A&A...266..117A, 2008ApJ...687..133I, 2015MNRAS.453.4102D, 2020MNRAS.492..468D}) and strong-line methods (e.g., \citealt{Storchi_Bergmann_1998, 10.1093/mnras/stx150, 2020MNRAS.492.5675C}). On the other hand, for Low-Ionization Nuclear Emission-line Regions (LINERs), chemical abundance studies are rarely found in the literature. LINERs appear in 1/3 of galaxies in the local universe \citep{netzer_2013} and their ionization sources are still an open problem in astronomy. \cite{1980A&A....87..152H} suggested that the main ionization/heating source of these nuclei are  gas shocks. Later, \cite{1983ApJ...269L..37H} and \cite{1983ApJ...264..105F} proposed that the accretion gas in a central black hole (AGN) should be responsible for the ionization of LINERs. Thus, the difference between LINERs and other AGN types would consist in the order of magnitude of the ionization parameter \citep{1993ApJ...417...63H}.  However, \cite{10.1093/mnras/213.4.841} and \cite{1992ApJ...399L..27S} proposed that the LINER-like emission is produced by photoionization due to hot stars that came out of the main sequence (e.g., in the post-Asymptotic Giant Branch, post-AGB). Based on this scenario, \cite{Taniguchi_2000} showed that photoionization models considering Planetary Nebula Nuclei (PNNs) with a temperature of $10^{5}$ K as ionizing sources, can reproduce the region occupied, at least, for a subset of type 2 LINERs in optical emission-line ratio diagnostic diagrams.  \cite{2014arXiv1409.2966W} found that these objects have composite ionizing sources, and more than one mechanism could be responsible for the  gas ionization. The same scheme was also proposed by \cite{Yan_2012}, \cite{2013A&A...558A..43S}, and \cite{2013A&A...558A..34B}. 

The unknown nature of the ionizing sources and excitation mechanisms of LINERs make it difficult to determine their metallicity  using the  $T_{\rm e}$-method and/or strong-line methods.  \citet{Storchi_Bergmann_1998} found that  their 
 calibrations work very well for the Seyfert galaxies, yielding abundance values that agree with those obtained from the extrapolation of O/H abundance gradients to the central regions of the host galaxies. However, for the LINERs, their calibrations yield lower values (up to $\sim 0.5$ dex) than those derived through  the extrapolation of O/H abundance gradients, and they concluded that their model  assumptions were not suitable for LINERs. Regarding the  $T_{\rm e}$-method application to LINERs,
\citet{2018MNRAS.481..476Y} determined the electron temperature in the S$^{+}$ and O$^{+}$ zones for  a sample of  quiescent red sequence galaxies with low ionization regions using spectra from the Sloan Digital Sky Survey (SDSS)  and compared the temperature-sensitive line ratios  with different model predictions to infer the metallicity of these galaxies. They found that neither the photoionization models simulated pure AGN nor the shock models simultaneously reproduced all studied line ratios.
 For example,  high temperatures estimated from $[\ion{S}{ii}]$ and $[\ion{N}{ii}]$ line ratios imply  subsolar metallicities, while their   $[\ion{N}{ii}]/[\ion{O}{ii}]$ line ratios required supersolar metallicities.  In a recent work, \cite{2021MNRAS.505.2087K} estimated the oxygen abundance of the LINER nucleus of the UGC\,4805 galaxy through the extrapolation of the radial abundance gradient, as well as strong emission-line calibrations for AGNs and photoionization models assuming gas accretion into a black hole (representing an AGN) and post-AGB stars with different temperatures. These authors found that all O/H abundance estimations agreed with each other. Although both AGN and post-AGB models were  able to reproduce the observational data, the high gas excitation level that must be maintained at kpc scales and the results from the WHAN diagram suggest that the main ionizing source of the UGC 4805 nucleus probably has a stellar origin rather than an AGN.

 In this paper, we propose two new metallicity abundance calibrations for LINERS by using  $N2 = \log$([\ion{N}{ii}]$\lambda$ 6584/H$\alpha)$ and  $O3N2 = \log \left(\frac{[\mathrm{OIII}]\lambda\,5007/ \mathrm{H}\beta}{[\mathrm{NII}]\lambda\,6583/\mathrm{H}\alpha}\right)$ strong-emission line indexes. To calibrate  the new relations, we combined observational data  with photoionization models assuming post-AGB stars as ionizing sources. Our  sample is composed of 43 galaxies with LINER emission  in their nuclear region and  with SF emission  in their disks.
 The observational data were compiled from the Mapping Nearby Galaxies at APO (MaNGA, \citealt{2015ApJ...798....7B}) survey.
 This paper is organized as follows:  Section~\ref{dataobs} describes the observational data   and the selection criteria of the sample. The methodology used to obtain the metallicity calibrations is presented in Section~\ref{methods}. Section~\ref{res} contains the results obtained, which are discussed in  Section~\ref{disc}. The conclusion of the outcome is provided in   Section~\ref{conc}.

%%%%%%%%%%%%%%%%%%%%%%%%%%%%%%%%%%%%%%%%%

\section{Observational Data}
\label{dataobs}

\subsection{MaNGA overview and measurements}

MaNGA is an Integral Field Spectroscopy (IFS) survey\footnote{\url{https://www.sdss.org/surveys/manga/}}
\citep{2015ApJ...798....7B}, which observed  about 10\,000 galaxies in the local universe.  This survey is part of the Sloan Digital Sky Survey (SDSS-IV, \citealt{2017AJ....154...28B}) and was performed using a 2.5 m telescope at the Apache Point Observatory.
The spectra have a wavelength coverage of 3\,600 -- 10\,300 \AA, with a spectral resolution of $R \sim$ 1\,400 
at $\lambda\sim$ 4\,000  \AA\  and $R \sim$ 2\,600 at $\lambda\sim$ 9\,000 \AA, with a spatial resolution of about 2.5" due to the mean local seeing (\citealt{2013AJ....146...32S, Drory_2015, 2017AJ....154...86W}).

MaNGA applies  a data analysis pipeline\footnote{\url{https://www.sdss.org/dr15/manga/manga-analysis-pipeline/}} (DAP) 
on the reduced IFS-cubes to produce 2D physical property maps \citep{belfiore19,westfall19}. Here, we describe, briefly, the DAP procedure.  First, it performs  a Voronoi re-binning of the cubes based on a  g-band weighted signal-to-noise ratio (S/N) image to reach a S/N of at least 10 on each target. Second, on the cube binned, the DAP fits the stellar continuum by using  the  Penalized PiXel-Fitting  (pPXF) method by  \cite{cappellari17}. The stellar templates are built by hierarchically clustering  the MILES stellar library \citep{sanchez-blazquez06,falcon-barroso11}. Finally, after 
the stellar continuum fitting continuum-subtracted spectra (the ``nebular'' ones), the DAP computes measurements of the emission line fluxes in two ways: one based on simple moments and  another based on a Gaussian fitting.  For further details, we refer the reader to \citet{belfiore19} and \citet{westfall19}. For this work, we use the 
2D maps built using the Gaussian fitting set taken from the Data Release15 (DR15) of MaNGA\footnote{\url{https://www.sdss.org/dr15/}} \citep{aguado19}.

\begin{figure*}
\includegraphics*[angle=0,width=0.32\textwidth]{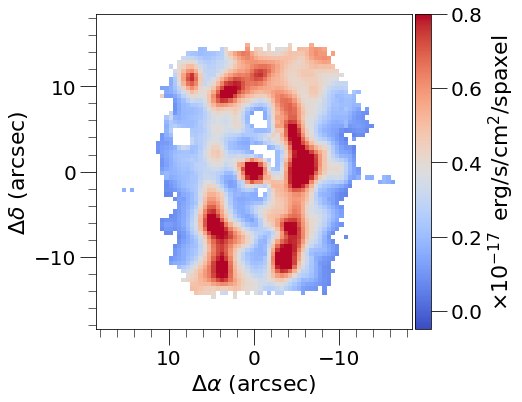}
\includegraphics*[angle=0,width=0.32\textwidth]{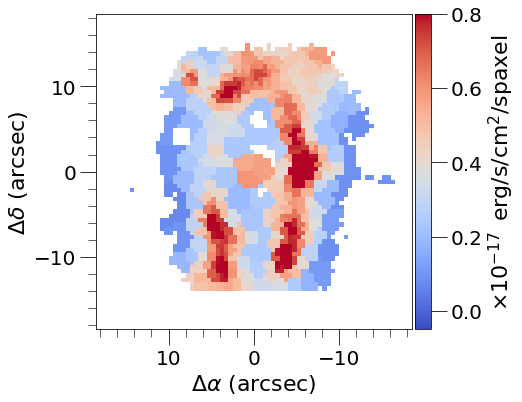}
\includegraphics*[angle=0,width=0.32\textwidth]{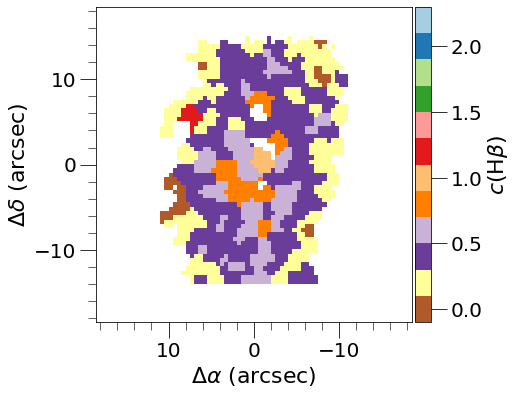}
\caption{Example of the tessellation binning based on the H$\beta$ S/N map for 8313-12705. Left: H$\beta$ emission-line flux map. Centre: Voronoi re-binned H$\beta$ emission-line flux map.
Right: Derived reddening map.}
\label{voronoi} 
\end{figure*}

To work with reliable data, we masked all spaxels with S/N $<3$ in the 2D emission-line fluxes and equivalent width maps. All emission-line intensities were reddening corrected using the extinction curve by \citet{cardelli89}. The theoretical value used for the  H$\alpha$/H$\beta$ ratio is  2.87, which was obtained for the recombination case B for an electron temperature of 10\,000 K at the limit of  the low density \citep{osterbrock06}. We derived the  2D reddening coefficient [$c$(H$\beta$)] maps re-binning by S/N the H$\beta$ image  using the publicly available Voronoi binning algorithm  by \citet{2003MNRAS.342..345C}. The target S/N per tessellation bin was 30 based.  Figure.~\ref{voronoi} is an example of  H$\beta$ flux and $c$(H$\beta$) tessellated maps for the galaxy 8313-12705.

\subsection{Sample selection}

As mentioned in \citet{2021MNRAS.505.2087K}, we  selected  objects with LINER emission in their nuclear regions and  SFs emission  in their disks. Furthermore, this sample is  restricted to objects with LINER emission with an integrated ${{\rm H}\alpha}$ equivalent width (EW$_{{\rm H}\alpha}$) lower  than 3 and higher than 0.5, which suggests that the ionization source of the nuclear region is probably post-AGBs. The sample selection followed the steps listed below.

\begin{enumerate}

\item  We used the log([\ion{O}{iii}]$\lambda$5007/\rm H$\beta$) vs.\ log([\ion{N}{ii}]$\lambda$6584/\rm H$\alpha$) diagnostic diagram proposed by \citet{1981PASP...93....5B},  called BPT, to  classify objects as \ion{H}{ii}-like regions, composite, and AGN-like objects considering the theoretical and  empirical criteria proposed by \citet{kewley01} and \citet{2003MNRAS.346.1055K}, respectively. The Seyfert and LINER objects are distinguished using the \citet{2006MNRAS.372..961K} criteria.

\begin{figure*}
\includegraphics*[angle=0,width=0.8\textwidth]{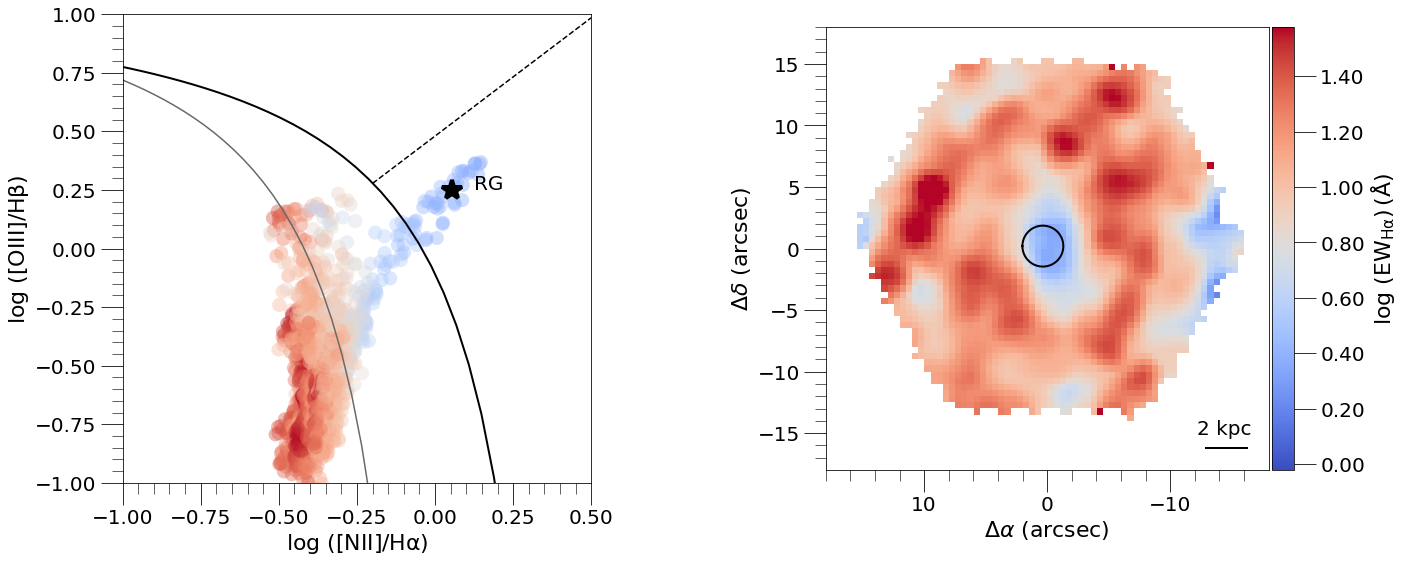}
\includegraphics*[angle=0,width=0.8\textwidth]{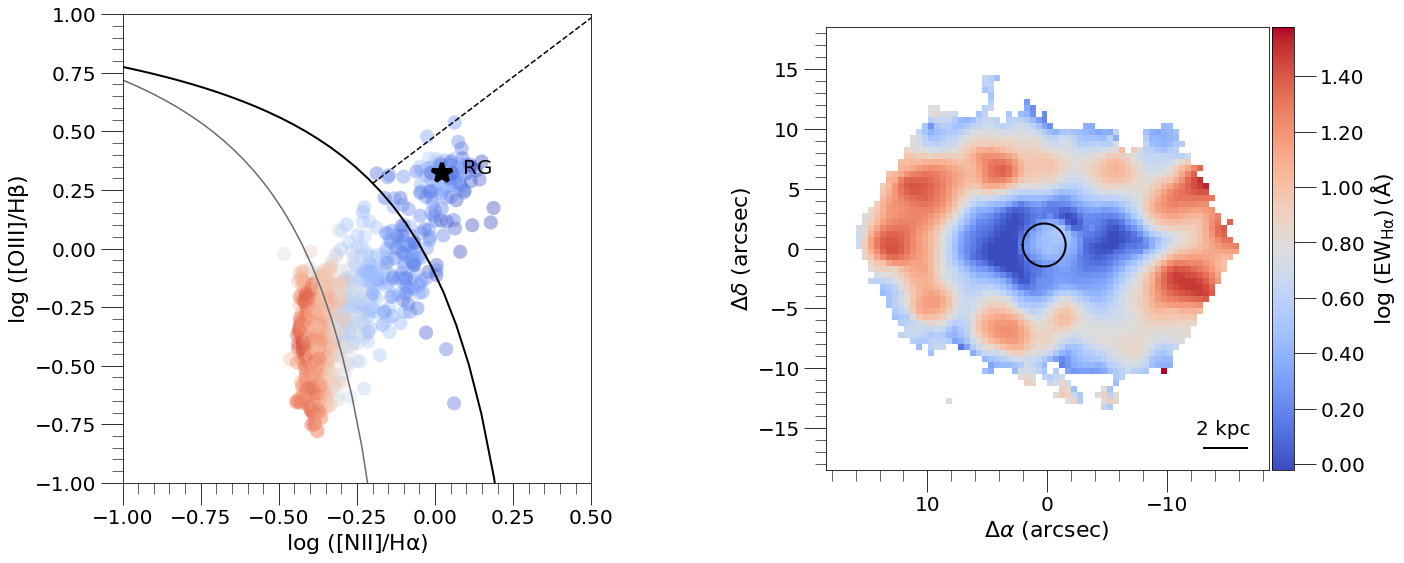}
\includegraphics*[angle=0,width=0.8\textwidth]{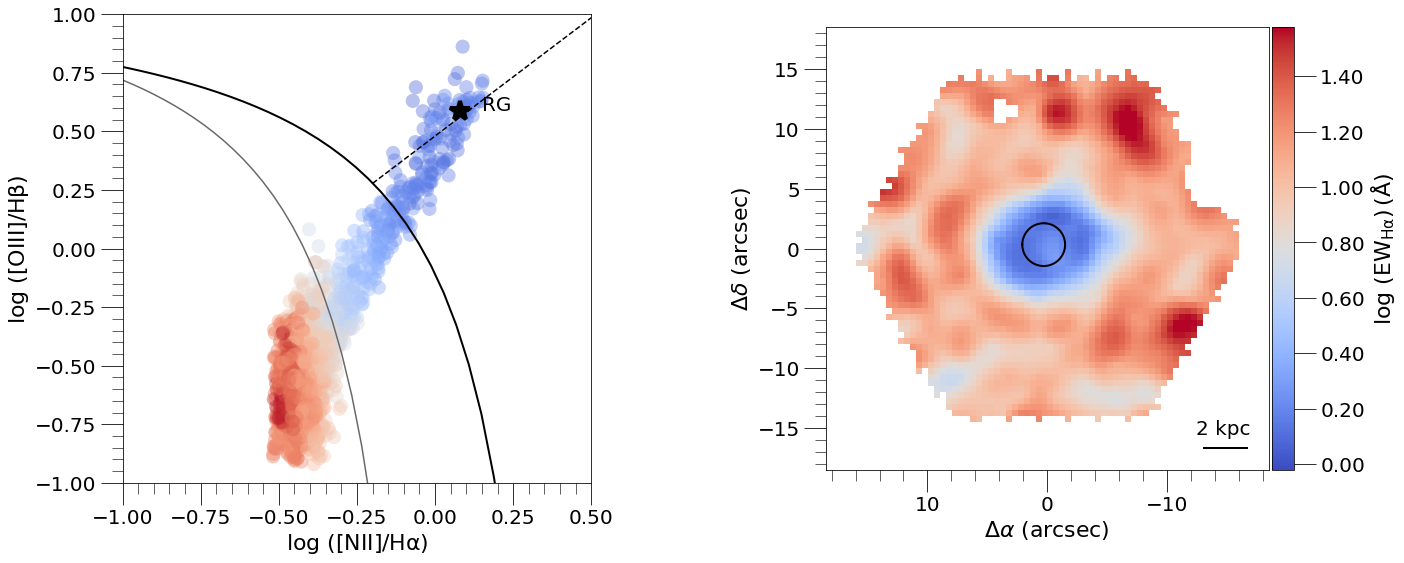}
\includegraphics*[angle=0,width=0.8\textwidth]{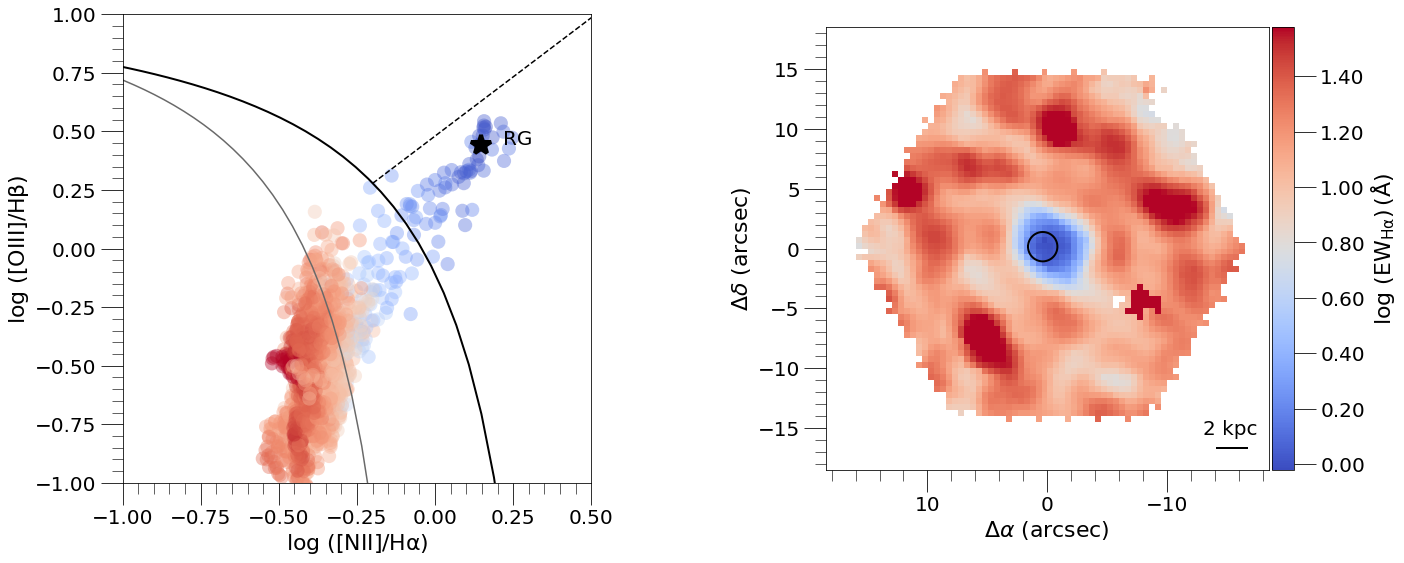}
\caption{ First column: log([\ion{O}{iii}]$\lambda$5007/\rm H$\beta$) vs.\ log([\ion{N}{ii}]$\lambda$6584/\rm H$\alpha$) diagnostic diagram , from top to bottom the galaxies shown are: 7495-12704, 7990-12704, 8249-12704, and 8318-12703. The black solid curve represents the theoretical upper limit for the star-forming regions proposed by \citet{kewley01} (ke01);  the grey solid curve is the  empirical star-forming limit proposed by \citet{2003MNRAS.346.1055K} (ka03); and the black dashed line represents the separation  between  Seyferts and LINERs \citep{2010MNRAS.403.1036C}. The region between the Ke01 and Ka03 lines is a denominated  composite region. Point colours are the same as in the second column and depend on the EW$_{H\alpha}$. Second column:  spatial distribution accordingly to the $\rm \log(EW_{H\alpha})$. The black stars represent the integrated estimations for the nuclear regions defined by the black circles.}
\label{maps}. 
\end{figure*}

\begin{figure*}
\includegraphics*[angle=0,width=0.8\textwidth]{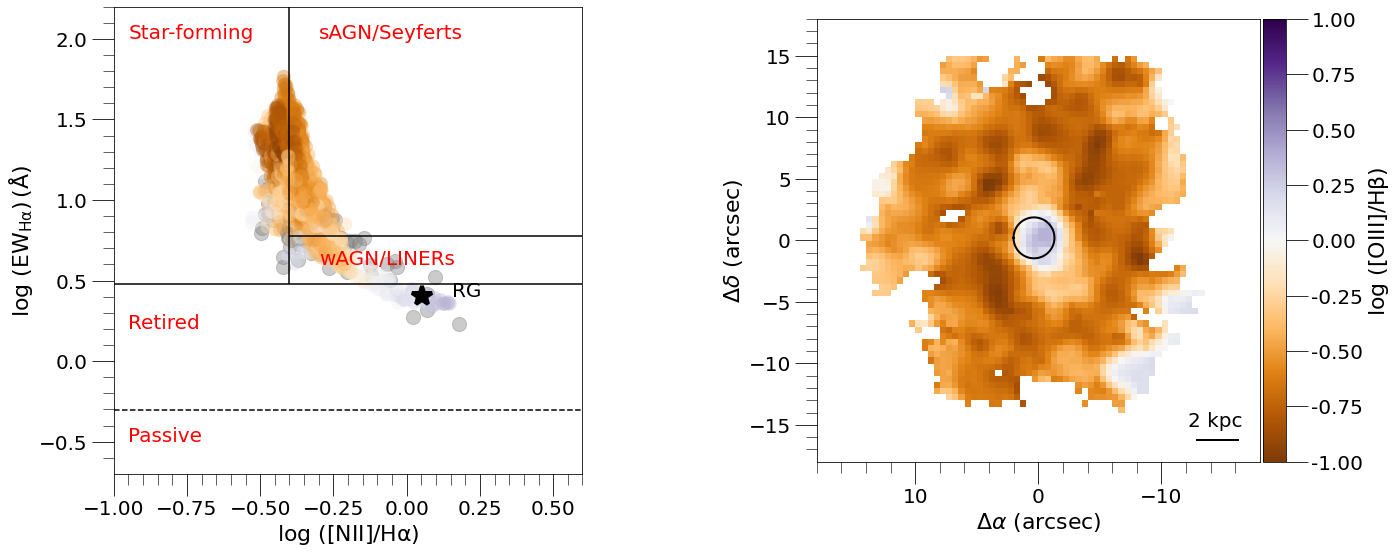}
\includegraphics*[angle=0,width=0.8\textwidth]{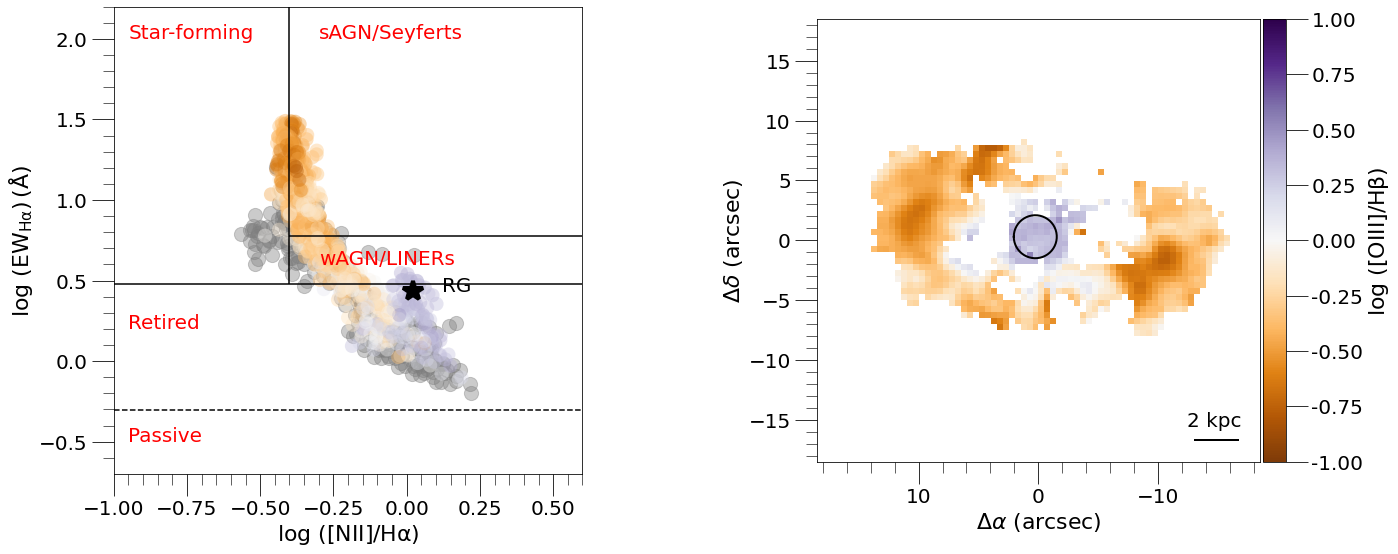}
\includegraphics*[angle=0,width=0.8\textwidth]{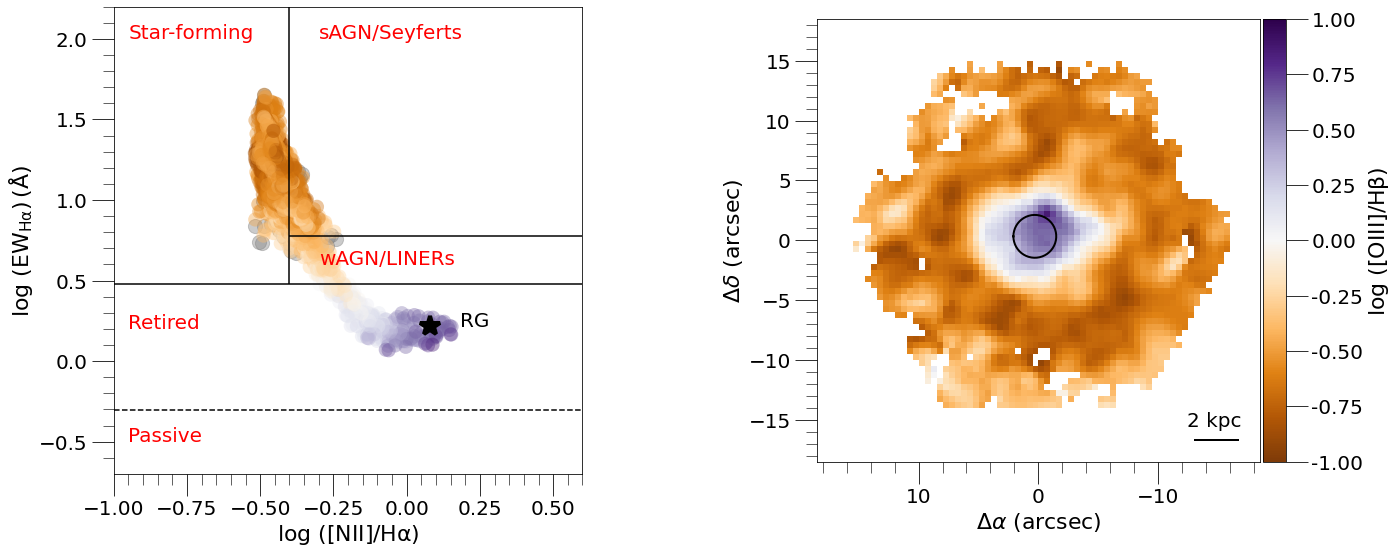}
\includegraphics*[angle=0,width=0.8\textwidth]{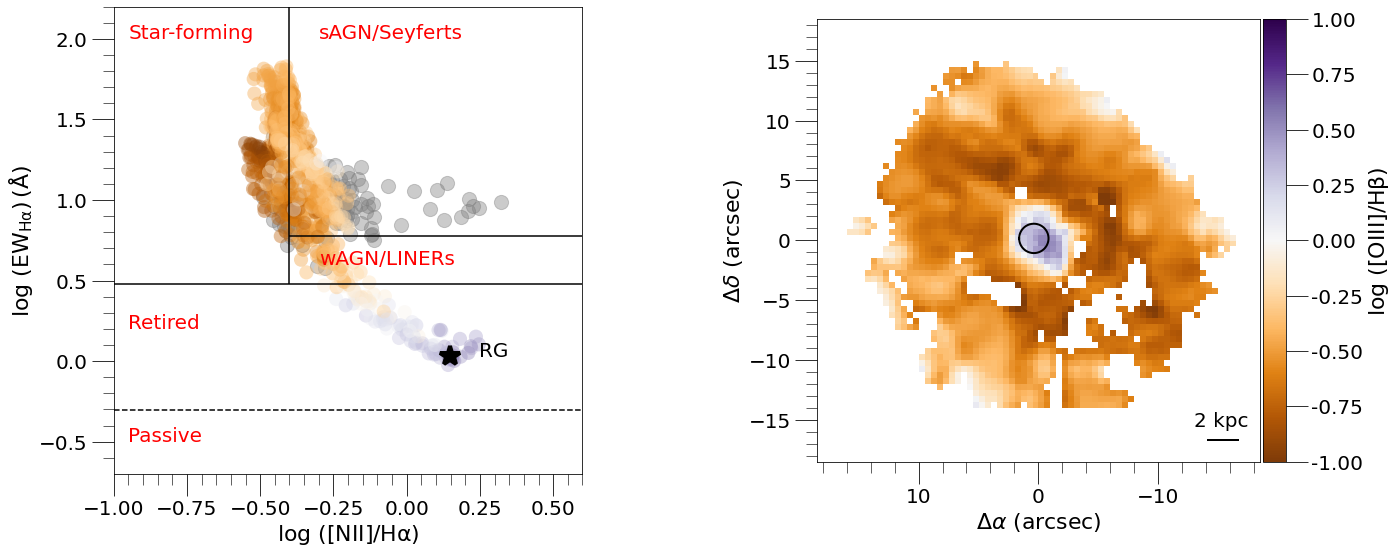}
\caption{ First column: examples of WHAN diagrams; from top to bottom the galaxies are: 7495-12704, 7990-12704, 8249-12704, and 8318-12703. Point colours are the same as in the second column and depend on the $\log([\ion{O}{iii}]$/H$\alpha$) ratio.  Second column: spatial distribution according to the  
$\log([\ion{O}{iii}]$/H$\alpha$) ratio.  The black stars represent the integrated estimations for the nuclear regions defined by the black circles.}
\label{maps1}. 
\end{figure*}

\begin{figure*}
\includegraphics*[angle=0,width=0.47\textwidth]{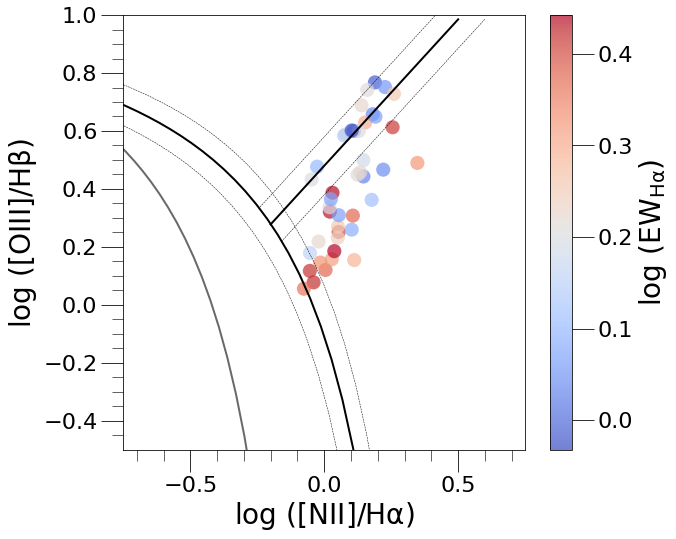}
\includegraphics*[angle=0,width=0.47\textwidth]{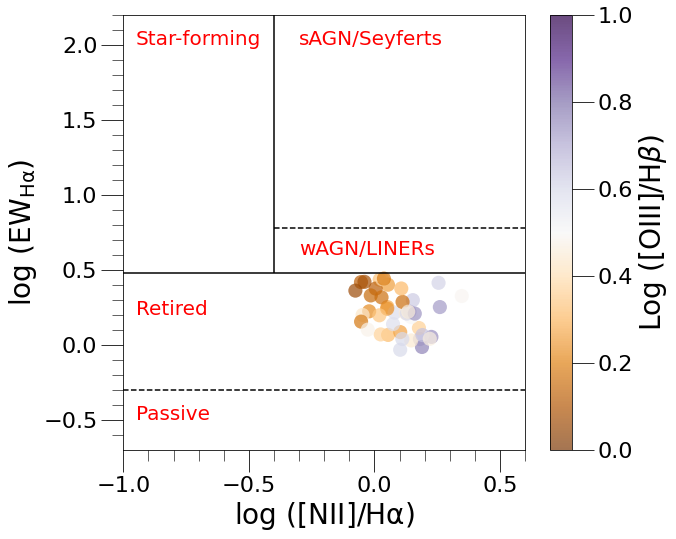}
\caption{BPT and WHAN diagrams as shown in Fig.~\ref{maps} and Fig.~\ref{maps1} but considering the nuclear-integrated flux for each galaxy in our sample.}
\label{bpt} 
\end{figure*}

\begin{figure}
\includegraphics*[angle=0,width=0.5\textwidth]{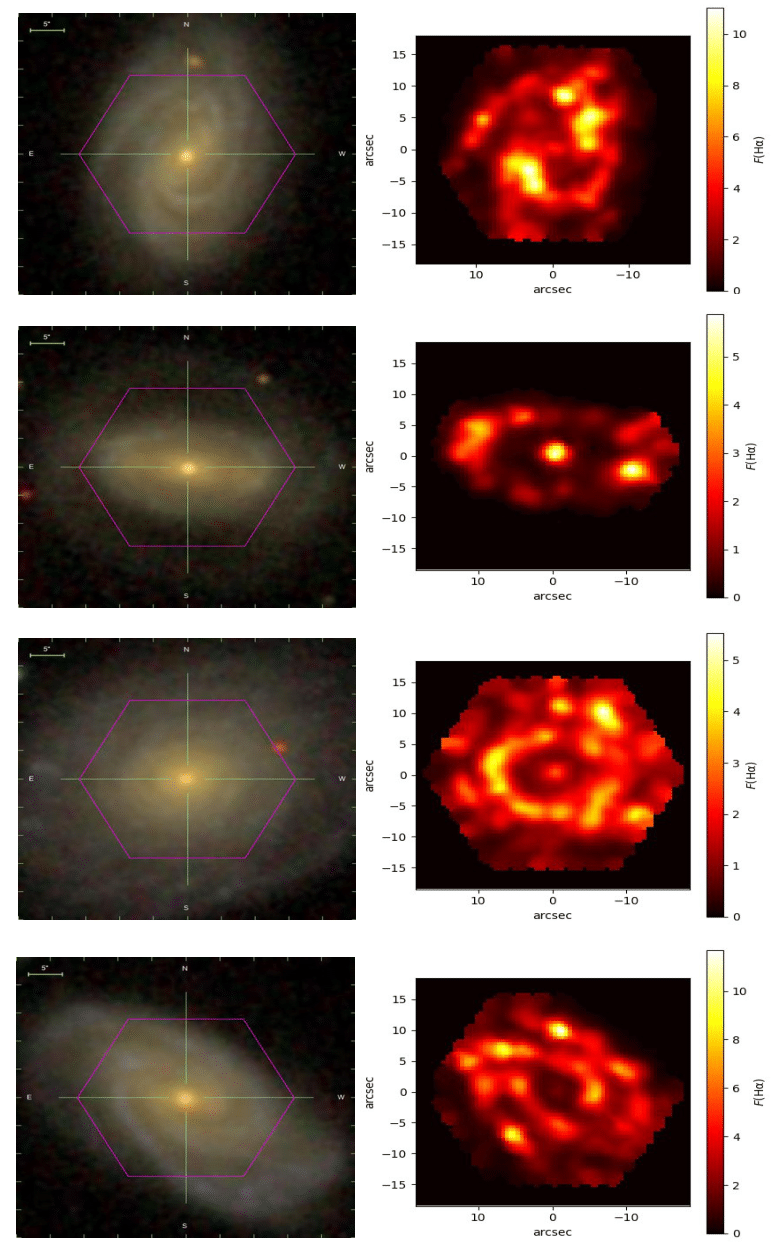}
\caption{Left panels: SDSS image  combining the \textit{gri} bands of  galaxies 7495-12704, 7990-12704, 8249-12704, and 8318-12703 (from top to bottom) with the MaNGA field of view  indicated by the purple hexagon. Right panels:  H$\alpha$ flux spatial distribution (in units of 10$^{-17}$ erg/cm$^2$/spaxel)}
\label{sample_ha}
\end{figure}

\item  We used the 
$\rm \log(EW_{H\alpha})$ vs.\ log([\ion{N}{ii}]$\lambda$6584/\rm H$\alpha$) diagnostic diagram, known as WHAN (\citealt{2011MNRAS.413.1687C}), to verify if the  nuclear regions initially classified  as LINER using  the BPT diagram occupy the  same area as in the WHAN diagram.
The WHAN diagram is  useful  to differentiate the nature of the ionization sources of LINERs, i.e.,  between evolved low-mass stars (like post-AGB stars) and low ionization AGNs. Using the $\rm EW_{H\alpha}$, this  diagram classifies objects into five classes of galaxies, namely:

\begin{enumerate}
  \item Pure star forming galaxies: $\log([\ion{N}{ii}]/\rm H\alpha) \: < \: -0.4$ and  $\rm EW_{H\alpha} \: > \: 3$ \AA.
  \item Strong AGNs: $\log({[\ion{N}{ii}]}/\rm H\alpha)\: > \: -0.4$ and  $\rm {EW_{H\alpha}} \: > \:  6$ \AA.
  \item Weak AGNs:  $\log({[\ion{N}{ii}]}/\rm H\alpha) \:  > \:  -0.4$ and  $\rm {EW_{H\alpha}}$ between 3 and 6 \AA.
  \item Retired Galaxies (RGs; i.e., fake AGNs):  $\rm {EW_{H\alpha}} \: < \: 3$ \AA. 
  \item Passive galaxies (actually, line-less galaxies): $\rm {EW_{H\alpha}}$ and $\rm {EW_{[\ion{N}{ii}]}} \: < \: 0.5$ \AA.
  
\end{enumerate}

From the BPT and WHAN diagrams together with their spatial distributions we selected all  objects whose nuclei are in the LINER zone in the first diagram and  the RG zone in the second. We also required that the disk spaxels be in the SF zone of both diagrams.   The BPT and WHAN diagrams are  
in Fig.~\ref{maps} and Fig.~\ref{maps1},  respectively, together with their spatial distributions, for  the galaxies 7495-12704, 7990-12704, 8249-12704, and 8318-12703 belonging to our final sample. 
At this step, we selected all objects with the same spatial distribution pattern, following the procedures below.

 \item  A circular aperture with a radius of 1\,kpc was defined for each galaxy selected earlier. The integrated flux for each emission line considered was obtained for all nuclei in our sample by summing together all fluxes in this region.  The  BPT  and WHAN diagrams were repeated for the nuclear regions of the galaxies and are shown  in  Fig.~\ref{bpt}. We also include in this figure, as a black star, the nuclear integrated line ratios in the corresponding diagrams for each object, as  shown in  Fig.~\ref{maps} and Fig.~\ref{maps1}. 
 Finally, we selected all objects whose integrated nuclear emission is located in the LINER and RG zones in the BPT and WHAN diagrams, respectively.
 For the BPT diagram (Fig.~\ref{bpt}), deviations  of 0.1 dex were considered for the lines that represent the separation between Seyferts/LINERs and AGNs-like/SFs. These variations were assumed because  
 according to \citet{kewley01}, the lines separating the different regions have errors in the order of 0.1 dex, which are from the modelling due to the assumptions made in the chemical abundances, chemical depletion factors,  slope of the initial mass function, and stellar atmosphere models \citep{kewley01}. 
%According to the WHAN diagram  the nuclear regions of the galaxies of our sample could be ionized by post-AGB stars. 

% according to \citet{2010MNRAS.403.1036C} and \citet{kewley01} models, respectively, These  variations represents typical errors in observational
 %emission-line ratio intensities of SFs 
 %(e.g. \citealt{denicolo, 2003ApJ...591..801K, 2008MNRAS.383..209H, 2020ApJ...893...96B}) and AGNs  (e.g. \citealt{1978ApJ...223...56K, 1992A&A...266..117A, 2015ApJS..217...12D}).
 %Furthermore, according to \citet{kewley01}, the lines separating the different regions present %errors also on the order of 0.1 dex.

The final sample consists of 43 galaxies with a LINER type nucleus. The stellar mass of the hosting galaxies is in the range of 
$9.97 \: \la \: \log(M_{*}/\mathrm{M_{\odot}}) \: \la \: 11.16$
and the redshift is in the range of $0.02 \: \la  \: z \: \la \: 0.07$. For each galaxy of our sample, Table~\ref{parameters} lists the identification number of the plate, coordinates, redshift, distance, and integrated stellar mass. All information was taken from the manga.Pipe3D\footnote{https://data.sdss.org/datamodel/files/MANGA\_PIPE3D} and manga.drpall\footnote{https://data.sdss.org/datamodel/files/MANGA\_SPECTRO\_REDUX}(see \citealt{2016RMxAA..52..171S}).
Fig.~\ref{sample_ha} is the RGB SDSS image (combining the \textit{gri} bands) of the galaxies 7495-12704, 7990-12704, 8249-12704, and 8318-12703 together with  the MaNGA field of view   and the 2D maps of the spatial distribution of  H$\alpha$ flux. In Table~\ref{fluxos}, the reddening corrected emission line intensities, the reddening coefficient [$c$(H$\beta$)], and  the equivalent widths of $\rm H\alpha$ for each LINER nucleus are listed.
\end{enumerate}

\begin{table}
\centering
\caption{Identification number of the plate, coordinates, redshift ($z$), distance (Mpc) and logarithm of the integrated stellar mass in units of solar masses $(\mathrm{M_{\odot}})$.}
\label{parameters}
\begin{tabular}{@{}lccccccc@{}}
\hline
Plate-IFU &
  RA &
  DEC &
  redshift &
  distance   & $\log(\mathrm{M_{*}})$
	\\
&  \multicolumn{1}{c}{(deg)} & \multicolumn{1}{c}{(deg)} & &\multicolumn{1}{c}{(Mpc)} &   
$(\mathrm{M_{\odot}})$\\	
	
\hline
7495-12704	&	205.4384	&	27.0048	&	0.0292	&	125.29	&	10.72	\\
7977-3704	&	332.7987	&	11.8007	&	0.0272	&	116.48	&	10.36	\\
7977-12703	&	333.2018	&	13.3341	&	0.0744	&	330.60	&	11.00	\\
7990-6103	&	261.2849	&	58.7647	&	0.0296	&	125.33	&	10.32	\\
7990-12704	&	262.4861	&	58.3974	&	0.0271	&	116.46	&	10.51	\\
8083-12704	&	50.6968	    &	0.1494	&	0.0231	&	98.90	&	10.44	\\
8131-9102	&	112.2214	&	41.3078	&	0.0586	&	256.17	&	9.97	\\
8140-12703	&	117.8985	&	42.8801	&	0.0323	&	138.58	&	10.83	\\
8243-9102	&	130.8217	&	52.7579	&	0.0591	&	260.66	&	11.10	\\
8243-12701	&	128.6877	&	52.7157	&	0.0452	&	196.78	&	11.16	\\
8247-3701	&	136.6714	&	41.3651	&	0.0252	&	107.68	&	10.35	\\
8249-12704	&	137.3775	&	45.9524	&	0.0271	&	116.46	&	10.67	\\
8252-12702	&	145.5309	&	48.1549	&	0.0339	&	143.10	&	10.88	\\
8254-3704	&	164.0822	&	43.7549	&	0.0362	&	156.36	&	10.76	\\
8257-1902	&	166.2978	&	46.1029	&	0.0371	&	160.81	&	10.64	\\
8258-12704	&	167.7765	&	43.6330	&	0.0253	&	107.68	&	10.79	\\
8259-9102	&	178.5399	&	44.3661	&	0.0620	&	274.51	&	10.99	\\
8313-9102	&	239.9880	&	41.4778	&	0.0335	&	143.05	&	10.72	\\
8313-12705	&	242.6825	&	41.1486	&	0.0319	&	134.22	&	10.91	\\
8318-12703	&	196.2324	&	47.5036	&	0.0396	&	169.85	&	10.99	\\
8320-9102	&	206.8303	&	21.8338	&	0.0527	&	228.69	&	11.02	\\
8332-12705	&	209.2520	&	43.3620	&	0.0333	&	143.03	&	10.93	\\
8330-9102	&	205.0114	&	40.4209	&	0.0245	&	103.31	&	10.45	\\
8332-6103	&	207.6574	&	43.7641	&	0.0489	&	210.51	&	10.58	\\
8440-12704	&	136.1423	&	41.3978	&	0.0274	&	116.49	&	10.54	\\
8481-1902	&	237.6539	&	53.3906	&	0.0654	&	288.53	&	10.77	\\
8482-12703	&	245.5031	&	49.5208	&	0.0500	&	215.07	&	10.96	\\
8549-3703	&	241.4164	&	46.8466	&	0.0575	&	251.55	&	10.67	\\
8550-6103	&	247.6387	&	39.8307	&	0.0249	&	103.36	&	10.41	\\
8550-12704	&	247.0584	&	40.3138	&	0.0334	&	143.04	&	10.72	\\
8550-12705	&	249.1357	&	39.0279	&	0.0303	&	129.72	&	11.04	\\
8552-9101	&	226.9119	&	44.5563	&	0.0664	&	293.17	&	10.86	\\
8601-12705	&	250.1231	&	39.2351	&	0.0300	&	129.68	&	10.47	\\
8588-9101	&	250.1562	&	39.2216	&	0.0355	&	151.95	&	10.63	\\
8138-3702	&	116.0979	&	44.5277	&	0.0500	&	215.07	&	10.72	\\
8138-9101	&	117.3026	&	45.5103	&	0.0535	&	233.21	&	10.84	\\
8482-3704	&	245.4124	&	49.4488	&	0.0328	&	138.65	&	10.86	\\
8482-9101	&	241.7996	&	48.5726	&	0.0437	&	187.85	&	10.84	\\
8554-1902	&	183.1133	&	35.8835	&	0.0231	&	98.90	&	10.02	\\
8603-12703	&	247.2827	&	40.6650	&	0.0303	&	129.71	&	10.49	\\
8604-12703	&	247.7642	&	39.8385	&	0.0309	&	129.79	&	10.79	\\
8604-6102	&	246.0735	&	39.2110	&	0.0303	&	129.72	&	10.68	\\
8606-3702	&	253.7939	&	36.9063	&	0.0239	&	98.97	&	10.30	\\

\hline
\end{tabular}
\end{table}

\begin{table*}
\centering
\caption{Reddening corrected emission-line intensities (in relation to H$\beta$=1.00) were derived for each LINER nucleus in our sample. Values of
the $\rm EW_{H\alpha}$ and the reddening coefficient [$c$(H$\beta$)] are also listed.}
\label{fluxos}
\begin{tabular}{@{}lcccccccccc@{}}
\hline
Plate-IFU  & [\ion{O}{II}]\,$\lambda$3727   & H$\beta$ & [\ion{O}{III}]\,$\lambda$5007 &  H$\alpha$     & [\ion{N}{II}]\,$\lambda$6584      	      & 	$ \rm EW_{H\alpha}$  &   $c$(H$\beta$)  \\     			     
\hline
7495-12704	&	1.77	$\pm$	0.03	&	1.00	$\pm$	0.01	&	1.79	$\pm$	0.01	&	2.87	$\pm$	0.01	&	3.25	$\pm$	0.01	&	2.52	$\pm$	0.05	&	0.36	\\
7977-3704	&	5.06	$\pm$	0.09	&	1.00	$\pm$	0.02	&	1.65	$\pm$	0.02	&	2.87	$\pm$	0.01	&	2.73	$\pm$	0.02	&	1.68	$\pm$	0.04	&	0.40	\\
7977-12703	&	3.93	$\pm$	0.24	&	1.00	$\pm$	0.06	&	1.31	$\pm$	0.06	&	2.87	$\pm$	0.03	&	2.54	$\pm$	0.03	&	2.62	$\pm$	0.08	&	0.31	\\
7990-6103	&	11.40	$\pm$	0.30	&	1.00	$\pm$	0.04	&	2.71	$\pm$	0.05	&	2.87	$\pm$	0.02	&	2.57	$\pm$	0.01	&	1.58	$\pm$	0.07	&	0.24	\\
7990-12704	&	4.35	$\pm$	0.04	&	1.00	$\pm$	0.01	&	2.10	$\pm$	0.01	&	2.87	$\pm$	0.01	&	3.01	$\pm$	0.01	&	2.72	$\pm$	0.06	&	0.28	\\
8083-12704	&	2.27	$\pm$	0.03	&	1.00	$\pm$	0.01	&	1.32	$\pm$	0.01	&	2.87	$\pm$	0.01	&	2.90	$\pm$	0.01	&	2.38	$\pm$	0.06	&	0.26	\\
8131-9102	&	25.56	$\pm$	0.91	&	1.00	$\pm$	0.09	&	4.55	$\pm$	0.12	&	2.87	$\pm$	0.04	&	4.35	$\pm$	0.01	&	1.09	$\pm$	0.05	&	0.33	\\
8140-12703	&	8.77	$\pm$	0.18	&	1.00	$\pm$	0.02	&	1.87	$\pm$	0.03	&	2.87	$\pm$	0.01	&	3.23	$\pm$	0.01	&	1.79	$\pm$	0.04	&	0.41	\\
8243-9102	&	10.61	$\pm$	0.35	&	1.00	$\pm$	0.04	&	3.09	$\pm$	0.04	&	2.87	$\pm$	0.02	&	6.39	$\pm$	0.01	&	2.12	$\pm$	0.04	&	0.54	\\
8243-12701	&	12.46	$\pm$	0.30	&	1.00	$\pm$	0.04	&	4.26	$\pm$	0.05	&	2.87	$\pm$	0.02	&	4.08	$\pm$	0.02	&	1.99	$\pm$	0.05	&	0.52	\\
8247-3701	&	3.58	$\pm$	0.03	&	1.00	$\pm$	0.01	&	1.44	$\pm$	0.01	&	2.87	$\pm$	0.01	&	3.06	$\pm$	0.01	&	2.08	$\pm$	0.03	&	0.29	\\
8249-12704	&	4.25	$\pm$	0.06	&	1.00	$\pm$	0.02	&	3.88	$\pm$	0.02	&	2.87	$\pm$	0.01	&	3.45	$\pm$	0.01	&	1.64	$\pm$	0.04	&	0.34	\\
8252-12702	&	27.75	$\pm$	0.53	&	1.00	$\pm$	0.06	&	5.86	$\pm$	0.06	&	2.87	$\pm$	0.02	&	4.44	$\pm$	0.01	&	0.97	$\pm$	0.03	&	0.40	\\
8254-3704	&	5.62	$\pm$	0.21	&	1.00	$\pm$	0.04	&	2.04	$\pm$	0.05	&	2.87	$\pm$	0.03	&	3.25	$\pm$	0.02	&	1.17	$\pm$	0.05	&	0.55	\\
8257-1902	&	3.90	$\pm$	0.06	&	1.00	$\pm$	0.01	&	1.40	$\pm$	0.02	&	2.87	$\pm$	0.01	&	2.77	$\pm$	0.01	&	2.14	$\pm$	0.03	&	0.45	\\
8258-12704	&	35.74	$\pm$	0.38	&	1.00	$\pm$	0.04	&	5.64	$\pm$	0.04	&	2.87	$\pm$	0.01	&	4.84	$\pm$	0.01	&	1.13	$\pm$	0.04	&	0.26	\\
8259-9102	&	8.45	$\pm$	0.29	&	1.00	$\pm$	0.05	&	2.44	$\pm$	0.06	&	2.87	$\pm$	0.02	&	3.08	$\pm$	0.02	&	2.74	$\pm$	0.06	&	0.42	\\
8313-9102	&	5.05	$\pm$	0.11	&	1.00	$\pm$	0.03	&	2.30	$\pm$	0.03	&	2.87	$\pm$	0.02	&	4.32	$\pm$	0.02	&	1.30	$\pm$	0.05	&	0.31	\\
8313-12705	&	5.55	$\pm$	0.09	&	1.00	$\pm$	0.02	&	3.16	$\pm$	0.03	&	2.87	$\pm$	0.02	&	4.02	$\pm$	0.02	&	1.53	$\pm$	0.05	&	0.38	\\
8318-12703	&	4.19	$\pm$	0.14	&	1.00	$\pm$	0.04	&	2.77	$\pm$	0.05	&	2.87	$\pm$	0.03	&	4.02	$\pm$	0.04	&	1.07	$\pm$	0.05	&	0.40	\\
8320-9102	&	4.70	$\pm$	0.16	&	1.00	$\pm$	0.04	&	2.81	$\pm$	0.05	&	2.87	$\pm$	0.03	&	3.82	$\pm$	0.03	&	1.63	$\pm$	0.06	&	0.36	\\
8332-12705	&	18.37	$\pm$	0.35	&	1.00	$\pm$	0.04	&	4.46	$\pm$	0.05	&	2.87	$\pm$	0.02	&	4.46	$\pm$	0.01	&	1.16	$\pm$	0.03	&	0.47	\\
8330-9102	&	2.29	$\pm$	0.07	&	1.00	$\pm$	0.03	&	1.82	$\pm$	0.03	&	2.87	$\pm$	0.02	&	3.63	$\pm$	0.03	&	1.22	$\pm$	0.06	&	0.48	\\
8332-6103	&	7.97	$\pm$	0.31	&	1.00	$\pm$	0.06	&	3.84	$\pm$	0.07	&	2.87	$\pm$	0.04	&	3.40	$\pm$	0.03	&	1.38	$\pm$	0.07	&	0.29	\\
8440-12704	&	3.82	$\pm$	0.04	&	1.00	$\pm$	0.01	&	1.71	$\pm$	0.01	&	2.87	$\pm$	0.01	&	3.23	$\pm$	0.01	&	1.74	$\pm$	0.04	&	0.23	\\
8481-1902	&	6.57	$\pm$	0.28	&	1.00	$\pm$	0.08	&	2.17	$\pm$	0.09	&	2.87	$\pm$	0.05	&	3.00	$\pm$	0.05	&	1.58	$\pm$	0.06	&	0.55	\\
8482-12703	&	10.04	$\pm$	0.24	&	1.00	$\pm$	0.05	&	4.10	$\pm$	0.05	&	2.87	$\pm$	0.03	&	5.17	$\pm$	0.03	&	2.59	$\pm$	0.10	&	0.30	\\
8549-3703	&	16.61	$\pm$	0.41	&	1.00	$\pm$	0.07	&	4.88	$\pm$	0.09	&	2.87	$\pm$	0.04	&	3.96	$\pm$	0.04	&	1.67	$\pm$	0.06	&	0.32	\\
8550-6103	&	4.73	$\pm$	0.04	&	1.00	$\pm$	0.01	&	1.53	$\pm$	0.01	&	2.87	$\pm$	0.01	&	3.13	$\pm$	0.01	&	2.77	$\pm$	0.05	&	0.38	\\
8550-12704	&	10.91	$\pm$	0.21	&	1.00	$\pm$	0.05	&	2.93	$\pm$	0.04	&	2.87	$\pm$	0.02	&	4.77	$\pm$	0.02	&	1.10	$\pm$	0.05	&	0.31	\\
8550-12705	&	9.88	$\pm$	0.24	&	1.00	$\pm$	0.04	&	3.98	$\pm$	0.04	&	2.87	$\pm$	0.02	&	3.71	$\pm$	0.02	&	1.10	$\pm$	0.04	&	0.68	\\
8552-9101	&	6.49	$\pm$	0.22	&	1.00	$\pm$	0.05	&	1.43	$\pm$	0.06	&	2.87	$\pm$	0.03	&	3.71	$\pm$	0.03	&	1.94	$\pm$	0.06	&	0.33	\\
8601-12705	&	3.48	$\pm$	0.17	&	1.00	$\pm$	0.03	&	1.51	$\pm$	0.04	&	2.87	$\pm$	0.02	&	2.54	$\pm$	0.02	&	1.43	$\pm$	0.06	&	0.55	\\
8588-9101	&	3.89	$\pm$	0.11	&	1.00	$\pm$	0.02	&	1.20	$\pm$	0.03	&	2.87	$\pm$	0.01	&	2.62	$\pm$	0.02	&	2.63	$\pm$	0.06	&	0.49	\\
8138-3702	&	13.32	$\pm$	0.26	&	1.00	$\pm$	0.05	&	3.98	$\pm$	0.05	&	2.87	$\pm$	0.03	&	3.86	$\pm$	0.01	&	1.62	$\pm$	0.05	&	0.74	\\
8138-9101	&	6.30	$\pm$	0.31	&	1.00	$\pm$	0.07	&	2.32	$\pm$	0.08	&	2.87	$\pm$	0.04	&	3.04	$\pm$	0.04	&	1.17	$\pm$	0.05	&	0.32	\\
8482-3704	&	10.00	$\pm$	0.13	&	1.00	$\pm$	0.02	&	2.03	$\pm$	0.02	&	2.87	$\pm$	0.01	&	3.67	$\pm$	0.01	&	2.38	$\pm$	0.04	&	0.52	\\
8482-9101	&	6.56	$\pm$	0.19	&	1.00	$\pm$	0.05	&	5.51	$\pm$	0.05	&	2.87	$\pm$	0.03	&	4.15	$\pm$	0.02	&	1.62	$\pm$	0.06	&	0.51	\\
8554-1902	&	4.16	$\pm$	0.05	&	1.00	$\pm$	0.02	&	3.00	$\pm$	0.02	&	2.87	$\pm$	0.02	&	2.70	$\pm$	0.01	&	1.26	$\pm$	0.06	&	0.29	\\
8603-12703	&	1.86	$\pm$	0.06	&	1.00	$\pm$	0.02	&	1.14	$\pm$	0.02	&	2.87	$\pm$	0.02	&	2.41	$\pm$	0.02	&	2.30	$\pm$	0.07	&	0.27	\\
8604-12703	&	12.39	$\pm$	0.15	&	1.00	$\pm$	0.03	&	5.35	$\pm$	0.03	&	2.87	$\pm$	0.01	&	5.23	$\pm$	0.02	&	1.79	$\pm$	0.05	&	0.48	\\
8604-6102	&	5.46	$\pm$	0.06	&	1.00	$\pm$	0.02	&	2.87	$\pm$	0.02	&	2.87	$\pm$	0.01	&	3.91	$\pm$	0.01	&	1.67	$\pm$	0.05	&	0.38	\\
8606-3702	&	8.84	$\pm$	0.12	&	1.00	$\pm$	0.03	&	4.00	$\pm$	0.03	&	2.87	$\pm$	0.02	&	3.63	$\pm$	0.01	&	0.93	$\pm$	0.05	&	0.48	\\

\hline
\end{tabular}
\end{table*}

%%%%%%%%%%%%%%%%%%%%%%%%%%%%%%%%%%%%%%%%%
\section{Methods}

\label{methods}

This section describes the methodology used to derive  two new oxygen abundance calibrations for LINERS by using the $N2$ and  $O3N2$ indexes. To calibrate  the new relations, we combined observational data  with photoionization models assuming post-AGB stars as ionizing sources.  In Section~\ref{modd}, we present the photoionization models used to derive oxygen abundances and ionization parameter values for the integrated fluxes from the central regions of the galaxies, and in  Section \ref{star_forming}, we discuss an indirect method, the determination of the nuclear oxygen abundances by extrapolating the metallicity gradients obtained through H{\sc ii} regions estimations, and we compare them with those derived using the new calibrations.

\subsection{Photoionization models}
\label{modd}

%to derive Seyfert 2 metallicities by \citet{10.1093/mnras/stx150} and \citet{2020MNRAS.492.5675C}  using optical  emission lines  and by \citet{2019MNRAS.486.5853D}  using  ultraviolet emission lines.  For each object in the sample, we estimated its metallicity   and ionization parameter ($U$) by interpolation between the nearest model results in the same way as performed by \citet{2021MNRAS.505.2087K}.

 We built photoionization model grids using the version $17.00$ of the    {\sc Cloudy} code   \citep{2017RMxAA..53..385F}. Post-AGB stars were considered as the ionization sources  since the nuclei of the objects in our sample have been classified as RGs according to the WHAN diagnostic diagram (see Fig.~\ref{bpt}).  These models are similar to the ones used by \citet{2021MNRAS.505.2087K}. A brief description of the input parameters is presented below.

% \sout{These authors showed that models with effective temperatures, $T_{\rm eff}$, of 50\,kK do not reproduce well the emission line ratios of the object analysed, therefore we do not take into account models with this effective temperature.}

\begin{enumerate}
    \item Spectral Energy Distribution (SED): we considered SED post-AGB star atmosphere models by \citet{2003A&A...403..709R} assuming effective temperatures ($T_{\rm eff}$) of   50, 100, and 190\,kK, with logarithm of surface gravity $\log$ (g) = 6.
     
    \item Metallicity: we considered the metallicity of the gas phase  ($Z$/Z$_{\odot}$) equal to 0.2, 0.5, 0.75, 1.0 and 2.0.  Assuming the solar oxygen abundance of 12 + $\log$ (O/H)$_\odot$ = 8.69 \citep{2001ApJ...556L..63A, 2009ARA&A..47..481A}, the corresponding  oxygen abundance range is of  $\rm 8.0 \: \la \: 12+log(O/H) \: \la \: 9.2$. 
   All  metals were linearly scaled with $Z$, with the exception of nitrogen, in which  we assumed the relation of   $\rm \log(N/O)=1.29\times [12 + \log(O/H)] - 11.84$ derived by \citet{2020MNRAS.492.5675C} from the abundance estimates of
   local SFs and Seyfert~2 nuclei.

    %\rojo{Are we sure that this relation derived for Seyfert 2 galaxies is the best choice for LINERs when we said that all of them are RGs?}
   %using  abundance estimations for type 2 AGNs and  \ion{H}{ii} regions and the abundances of heavy elements were scaled linearly with the metallicity.\\
 
    \item Electron Density ($N_{\rm e}$): we assumed three different  electron density values:  $N_{\rm e}$ = 100, 500 and 3\,000 $\rm cm^{-3}$, constant along the nebular radius. %These electron density values almost cover  our observational $N_{\rm e}$ estimations for our sample of galaxies, which are in the range of 
   % $35 \leq N_e \leq 3\,150\,  \rm cm^{-3}$, as derived using the  
    % $[\ion{S}{ii}] \,\lambda\, 6716/ [\ion{S}{ii}]\lambda\, 6731$ emission-line ratio and using the {\sc IRAF/TEMDEN} task,
    % assuming an electron temperature as $T_{\rm e} = 10\,000$ K.

    % \rojo{We have to say something about how we performed these estimations. Did we use PyNeb? We have to also include the Ne values in the "new" table.}

    \item Ionization Parameter: this parameter is defined as 
\begin{equation}
\label{elogu}
U = \frac{Q({\rm H)}}{4 \, \pi \, R_{{\rm 0}}^2 \, n(\rm H) \, \rm  c},
\end{equation}
 where $Q(\rm H)$ [$\rm s^{-1}$] is the number of hydrogen-ionizing photons emitted by the central ionizing object per second, $R_{0}$ [cm] is the distance from the ionization source to the inner surface of the ionized gas cloud, $n(\rm H)$ [cm$^{-3}$] is the total hydrogen density (ionized, neutral and molecular),  and $\rm c$ [cm\,s$^{-1}$] is the speed of light. We assumed the logarithm of $U$ in the range of $-4.0 \le \log U \le  -0.5$, with a step of 0.5 dex, which is  the same range of values assumed
by \citet{2021MNRAS.505.2087K}.

{\sc Cloudy} is a unidimensional code that assumes a central ionization source, which cannot represent the real situation in gaseous nebulae. In  most cases, a central ionization source cannot genuinely represent the situation, for example, for giant star-forming regions, since the stars may be spread out throughout the region (e.g., \citealt{2011MNRAS.413.2242M}). \citet{2009Ap&SS.324..199E} and \citet{2008A&A...482..209J} showed
that the distribution  of the hot stars (e.g. O-B stars) in relation to the gas alters the ionization structure and the electron temperature (also see \citealt{2022ApJ...927...37J}).  Hence, the ionization parameter partially depends on the spatial distribution of the ionizing sources with respect to the gas.  In our cases, we considered the integrated spectra of the nuclei to try to minimize the stellar distribution effect on the emergent spectra. The assumption of a single star with a representative effective temperature as the main ionizing source, as assumed in our post-AGB models, is a good approximation \citep[see, e.g.,][]{2019MNRAS.483.1901Z}, since in the case of giant  \ion{H}{ii} regions ionized by stellar clusters (e.g. \citealt{1996AJ....111.1252M, 2001A&A...380..137B}),  the hottest stars dominate the gas ionization \citep{2017MNRAS.466..726D}.

\end{enumerate}

 Once the photoionization models were built, diagrams of  $O3O2= \log$([\ion{O}{III}]$\lambda 5007$/[\ion{O}{II}]$\lambda 3727$)  versus  $N2$=log([\ion{N}{ii}] $\lambda$6584/H$\alpha$)  and  $O3N2= \log \left(\frac{[\mathrm{OIII}]\lambda\,5007/ \mathrm{H}\beta}{[\mathrm{NII}]\lambda\,6583/\mathrm{H}\alpha}\right)$  indexes for both the observational data and photoionization model results were obtained. Then,   the oxygen abundance (O/H) and the ionization parameter ($U$) of the gas phase of each LINER nucleus of the sample were determined by linear interpolation between the model results whose predict the nearest values to the measured emission line ratios.
 The methodology applied here is similar to  the one adopted by \citet{2021MNRAS.505.2087K}.
 
Note that  $O3O2$, $N2$, and  $O3N2$ indexes were selected in this work because the  former ratio is sensitive to the ionization degree of the gas phase, while the other two ratios mainly depend  on its metallicity. 
The $N2$ index was studied  as the
metallicity indicator for SFs by \citet{1994ApJ...429..572S}
and for Seyfert~2s by \citet{2020MNRAS.492.5675C}. 
The $O3N2$ index was first introduced by 
\citet{1979A&A....78..200A} as a  metallicity indicator for SFs,
and from there many calibrations have been  proposed in the literature (e.g.,\  \citealt{2004MNRAS.348L..59P,2009MNRAS.398..949P,marino}). 
The $N2$ ratio involves emission lines with close wavelengths, that makes it independent on the reddening correction and the uncertainties on the flux calibration  in contrast to the $O3N2$ index.
Moreover,  $N2$ involves ions with similar  ionization potentials, therefore, it is less dependent on   
the  ionization parameter in comparison to $O3N2$.

% to estimate the oxygen abundance (O/H) and the ionization parameter ($U$) of the gas phase of each object of the sample by linear interpolation of the model results
%which predict the nearest values of observational emission line ratios. We consider diagrams containing  the $O3O2= \log$([\ion{O}{III}]$\lambda 5007$/[\ion{O}{II}]$\lambda 3727$)  versus $N2$ and  $O3N2$.

\subsection{Extrapolating the metallicity gradient}
%\subsection{Extrapolated central intersect
%abundance }
\label{star_forming}

To verify  the validity of our calibrations, we compared  the oxygen abundance derived from  these  with measurements obtained from an independent method, which was the extrapolation of the radial oxygen
abundance gradient, obtained from \ion{H}{ii} region estimates along the galaxy disc to the central part of the host galaxies. 
This indirect method has been widely used in the literature \cite[e.g., ][]{1992MNRAS.259..121V,1998AJ....116.2805V,2004A&A...425..849P,Pilyugin+07, 2019MNRAS.483.1901Z, 2021MNRAS.505.2087K, 2022MNRAS.513..807D} 
and produces an independent estimation of the nuclear metallicity.

\begin{figure*}
\includegraphics*[angle=0,width=0.32\textwidth]{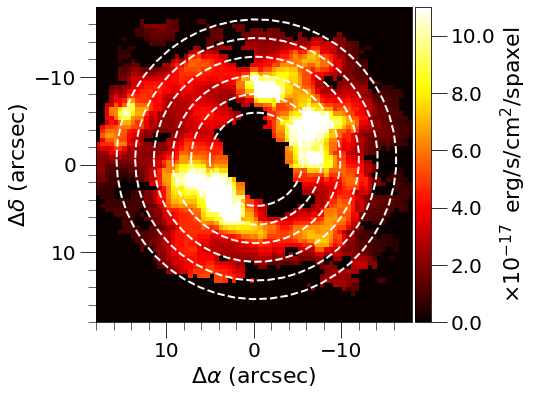}
\includegraphics*[angle=0,width=0.32\textwidth]{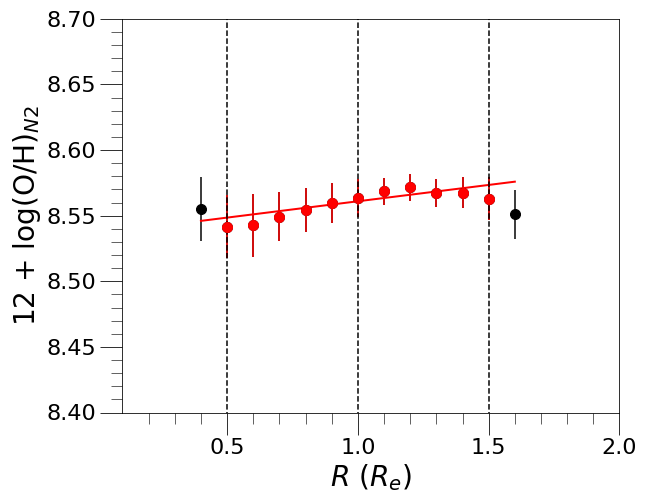}
\includegraphics*[angle=0,width=0.32\textwidth]{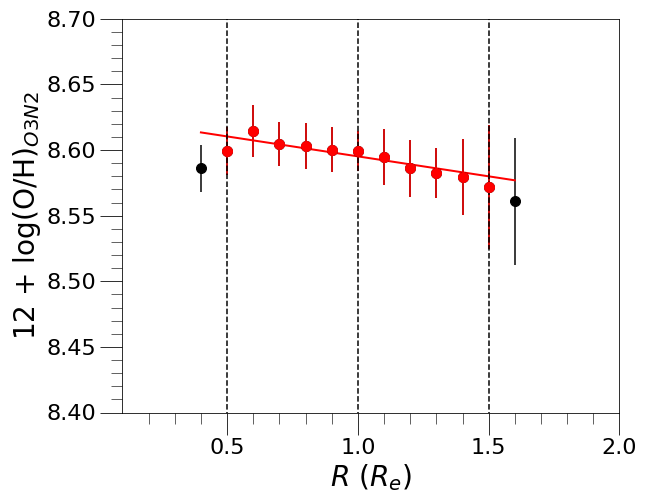}

\includegraphics*[angle=0,width=0.32\textwidth]{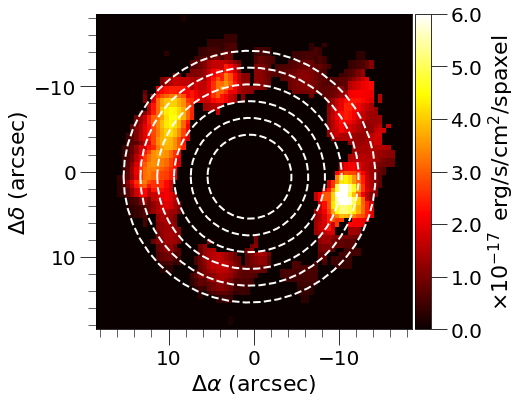}
\includegraphics*[angle=0,width=0.32\textwidth]{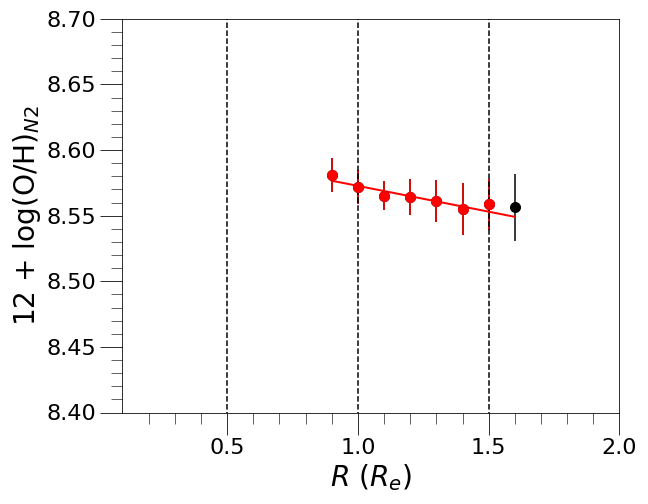}
\includegraphics*[angle=0,width=0.32\textwidth]{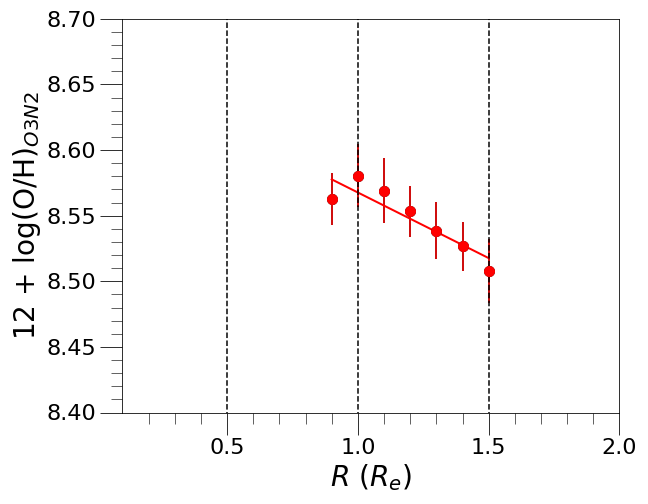}

\includegraphics*[angle=0,width=0.32\textwidth]{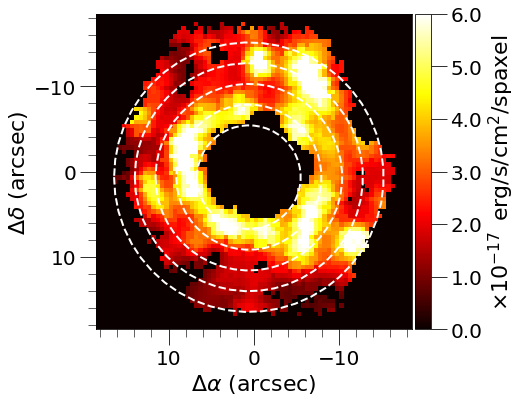}
\includegraphics*[angle=0,width=0.32\textwidth]{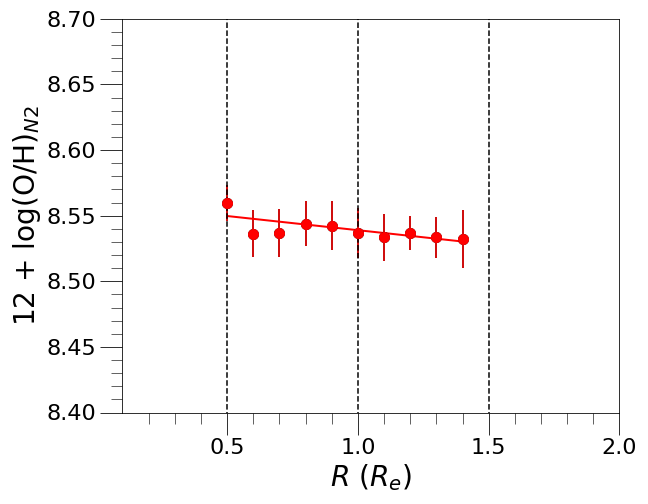}
\includegraphics*[angle=0,width=0.32\textwidth]{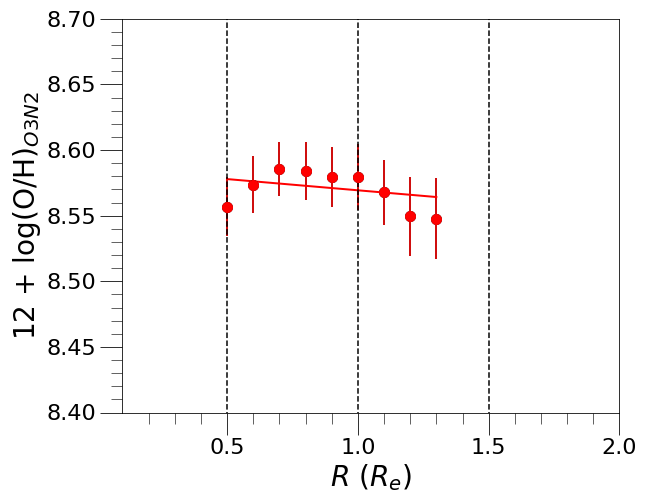}

\includegraphics*[angle=0,width=0.33\textwidth]{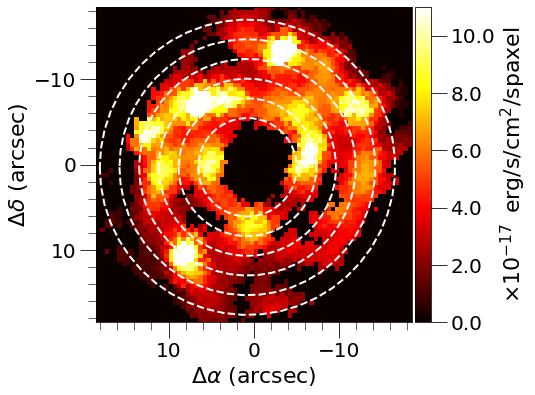}
\includegraphics*[angle=0,width=0.32\textwidth]{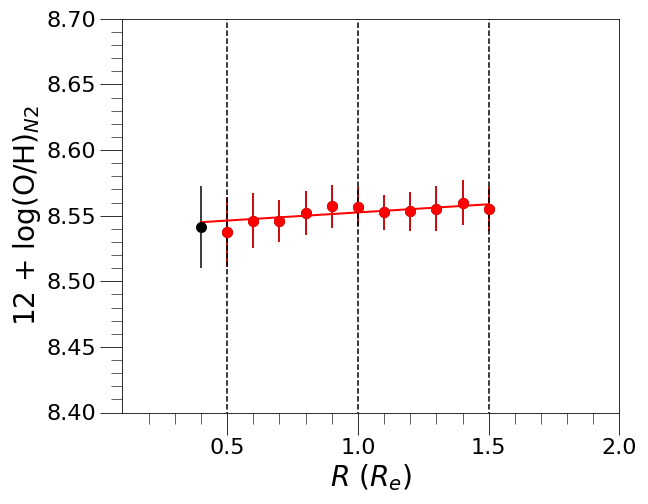}
\includegraphics*[angle=0,width=0.32\textwidth]{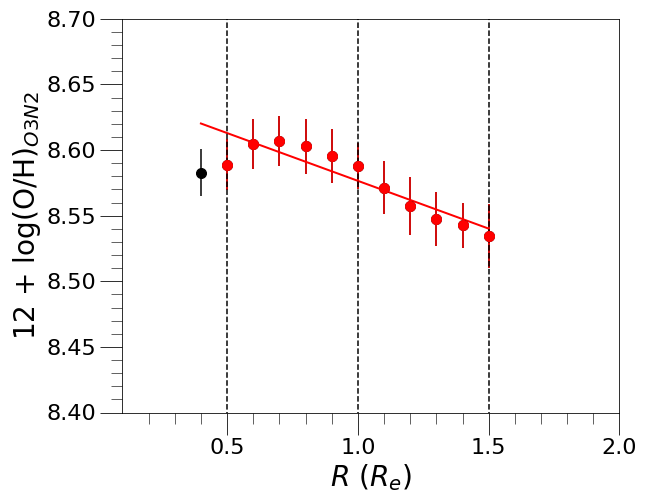}
\caption{The deprojected H$\alpha$ maps (left panels) and the metallicity gradients derived using $N2$ and $O3N2$ indexes (centre and right panels, respectively) for 7495-12704, 7990-12704, 8249-12704, and 8318-12703 (from top to bottom). The dashed white circles are  examples of the radial bins used to compute the metallicity profile from 0.5 and 1.5\,$R{\rm e}$, spaced in  0.2 $R{\rm e}$. The red lines are the best fitting between 0.5 and 1.5\,$R{\rm e}$ and the red open circles are the data points taken into account to fit within this range. The black circles are the data outside this range which are not considered.}
\label{fig_dep} 
\end{figure*}

Due to the random inclination of the galaxies in the sky, the projected galactic discs are
ellipticals. The radial profiles are calculated in elliptical annulus, producing some numerical artifacts. For example, the PSF beam-smearing  affects the inner radii, creating a spurious flatting of the metallicity gradient at the central regions \citep[e.g.,][]{belfiore17}, which is worse for higher inclinations. In view of that, we calculate the metallicity gradients between  0.5 and 1.5\,$R_{\rm e}$, where $R_{\rm e}$ is the half-light radius of each galaxy,
 using a radial bin of 0.1\,$R_{\rm e}$, and only computed the radial bin with at  least 20\% of the valid spaxels belonging to the corresponding annulus.
 Another important effect  is that the annulus aperture is oversampled  along the position angle of the galactic disk, while it is sub-sampled along the minor axis, yielding  biased statistics inside the apertures (e.g., average and median values). To correct this, we applied a novel method introduced by Hernandez-Jimenez (in prep.) to obtain an unbiased metallicity gradient. The key step is to deproject the 2D flux maps before  calculating the radial profile. To do this, we followed the recipe used by
 \citet{elmegreen92}, in which the image pixels are stretched along the minor-axis by a factor of the inverse cosine of the inclination angle\footnote{Taken from the manga.drpall table.}. This operation was
 performed by using the {\sc iraf imlintran} task. We set the task to preserve the total flux of the image.
 In the left panels of Fig.~\ref{fig_dep},
 we show the deprojected  2D  H$\alpha$ maps for 7495-12704, 7990-12704, 8249-12704, and 8318-12703. In these maps,  the central parts are ``removed'' because the ionizing sources  are not young stars. The elliptical-like SF rings (see Fig.~\ref{sample_ha}) observed in these galaxies are now circular due to the deprojection.
 
 Once the 2D emission line flux maps are deprojected, we built the 2D $O3N2$ and $N2$ index maps. Afterward, we converted them into 2D metallicity abundance  maps
 using the semi-empirical calibrations of these indexes proposed for SFs  by \cite{marino}.  These authors performed these calibrations using  observations of 3\,423 \ion{H}{II} regions from the CALIFA survey based on  the $T_{\rm e}$-method. The   $O3N2$  and $N2$ calibrations are  respectively given by 
 
\begin{equation}
 12 + \mathrm{\log(O/H)}  = 8.533 (\pm 0.012) - 0.214 (\pm 0.012) \times O3N2,
 \label{eqO3N2}
\end{equation}
 and 
\begin{equation}
 12 + \mathrm{\log(O/H)}  = 8.743 (\pm 0.027) +0.462 (\pm 0.024) \times N2,
 \label{eqN2} 
\end{equation}
which are valid  in the  $-1.1 \: < \:  O3N2 \: < \: 1.7$ range for equation  \ref{eqO3N2} and  in the $-1.6 \: < \:  N2 \: < \: -0.2$ range for equation \ref{eqN2}. Finally, we calculate the radial profile from H$\alpha$ flux-weighted median value of the abundance map along the circular annuli. Then,  the following linear fitting was obtained

%\begin{figure*}
%%\includegraphics*[angle=0,width=0.45\textwidth]{bpt.png}
%\includegraphics*[angle=0,width=0.47\textwidth]{map_hb_dep_8313-12705}
%\includegraphics*[angle=0,width=0.47\textwidth]{grad_N2_1_5_8313-12705_dep.png}
%\caption{Right: the deprojected H$\beta$ map of 8313-12705. The dashed black circles are the %radial bins used to compute the metallicity profile, from
%0.5 and 1.5\,$R{\rm e}$, spaced 0.1 $R{\rm e}$. Left:  The metallicity $N2$ index radial profile of 8313-12705. The red line is the best fitting between 0.5 and 1.5\,$R{\rm e}$. The red open circles are the data points taken into account to the fitting, while data as black circles are not considered.}
%\label{fig_dep} 
%\end{figure*}

% The deprojected galactocentric distances of  \ion{H}{ii} regions
%were calculated  to derive the radial distribution of abundances to   remove the effects of the inclination in the determinations  of the oxygen abundance.  The inclination angles were estimated by fitting ellipses to the isophotes of the galaxies in the XX band. This fitting was performed using  the {\sc ISOPHOTE} task of {\sc IRAF/STSDAS} from    \citet{1987MNRAS.226..747J}. The fitting is parametrized using values of position angle, ellipticity and  coordinates of the centre. 

%\jose{To derive the radial oxygen abundance of the \ion{H}{II} regions across the galaxy disks of our sample}, we %\cite{marino} \sout{to derive the oxygen abundance of the \ion{H}{II} regions across the galaxy disks of our sample}. \cite{marino} determined these calibrations using  observations of 3\,423 \ion{H}{II} regions from CALIFA survey based on  the $T_{\rm e}$-method. The   $O3N2$  and $N2$ calibrations are given respectively by 
%\begin{equation}
% 12 + \mathrm{\log(O/H)}  = 8.533 (\pm 0.012) - 0.214 (\pm 0.012) \times O3N2,
% \label{eqO3N2}
%\end{equation}
% and 
%\begin{equation}
% 12 + \mathrm{\log(O/H)}  = 8.743 (\pm 0.027) +0.462 (\pm 0.024) \times N2,
% \label{eqN2}
%\end{equation}
%and are valid  in the  $-1.1 \: < \:  O3N2 \: < \: 1.7$ range for equation  \ref{eqO3N2} and  in the $-1.6 \: < \:  %N2 \: < \: -0.2$ range for equation \ref{eqN2}. 

%The central oxygen abundance 
% $\rm 12 + \log(O/H)_{0}$  of the galaxies were obtained by extrapolating \sout{to} \jose{toward} the centre of the galaxy the linear fit:

\begin{equation}
\label{lingr}
{\rm  12 + \log (O/H)=12 +\log(O/H)_{0}}+ (grad \: \times R/R{\rm e}),
\end{equation}
where $\rm 12 + \log (O/H)$ is the oxygen abundance at a given
galactocentric  distance $R$ (in units of arcsec),  $grad$ is the regression slope, and $\rm 12 + \log(O/H)_{0}$  is the extrapolated value of the metallicity gradient at the centre of the galaxy ($R=0$ kpc). This fitting was performed for both $N2$ and $O3N2$ indexes obtaining two pairs of 12+log(O/H)$_0$ and grad for each galaxy.
We  computed  these metallicity gradients for  galaxies with at least 5 radial bins between 0.5 and 1.5\,$Re$.
In this way, we were able to perform the radial fitting for 33 and 32 galaxies  using the $N2$ and $O3N2$ indexes, respectively. The central and right panels of Fig.~\ref{fig_dep} present  for 7495-12704, 7990-12704, 8249-12704, and 8318-12703 the metallicity gradients derived from $N2$ and  $O3N2$ indexes, respectively. 

%%%%%%%%%%%%%%%%%%%%%%%%%%%%%%%%%%%%%%%%

\section{Results}
\label{res}

\begin{figure*}
\includegraphics*[angle=0,width=1\textwidth]{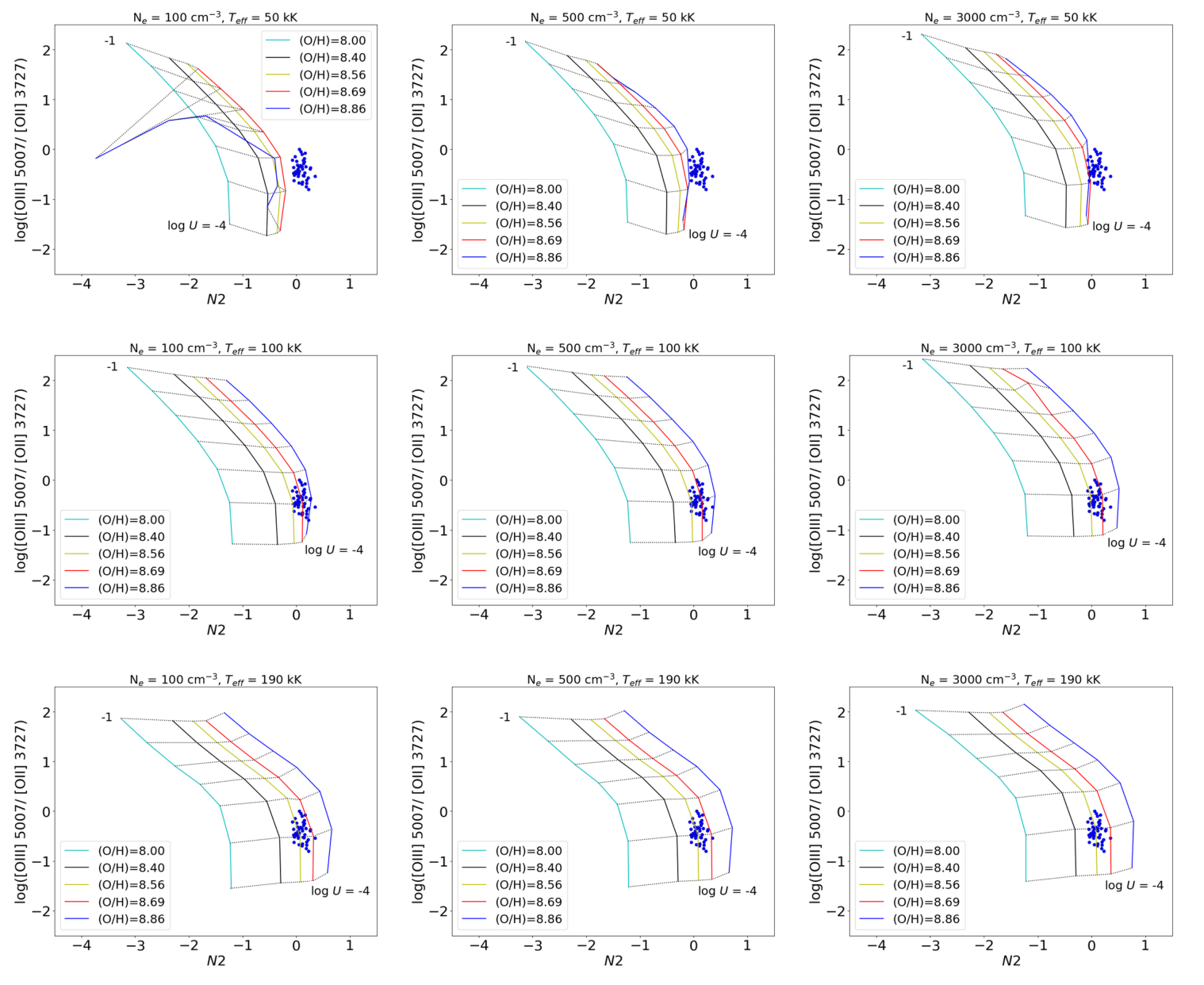}
\caption{$\log$([\ion{O}{III}]$\lambda 5007$/[\ion{O}{II}] $\lambda 3727$) versus $\log$([\ion{N}{II}]$\lambda 6584$/H$\alpha$) diagnostic diagram. Coloured solid lines connect the photoionization model results (see Section \ref{modd}) with the same oxygen abundance (O/H) and dotted line  models with the same ionization parameter ($U$), as indicated. The blue point represents the observational line ratios for each nucleus of our sample (see Section \ref{dataobs}).}
\label{grid_pagb_n2}
\end{figure*}

\begin{figure*}
\includegraphics*[angle=0,width=1\textwidth]{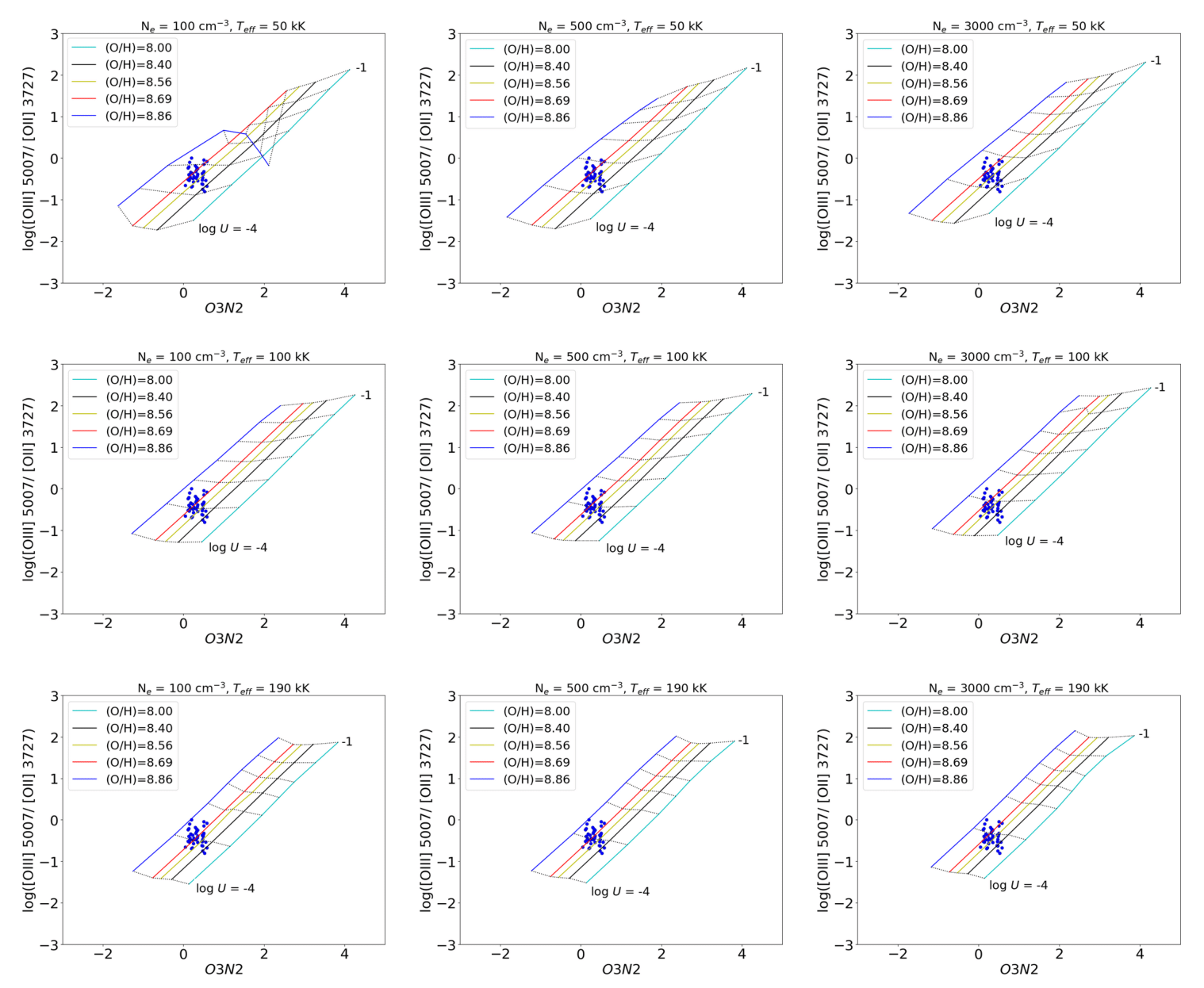}
\caption{Same as Fig.~\ref{grid_pagb_n2} but considering $O3N2$.}
\label{grid_pagb_o3n2}
\end{figure*}

The diagnostic diagrams 
containing the photoionization model results  and the observational data of the nuclei for the objects of our sample are shown in Figs.~\ref{grid_pagb_n2} and \ref{grid_pagb_o3n2}, for   $\log (\rm{[\ion{O}{iii}] \lambda 5007/[\ion{O}{ii}]\lambda3727})$ versus $N2$  index   and  $\log (\rm{[\ion{O}{iii}] \lambda 5007/[\ion{O}{ii}]\lambda3727})$ versus  the  $O3N2$ indexes, respectively.   Nine model grids with different  $N_{\rm e}$ and $T_{\rm eff}$ values  are considered  for the $N2$ and $O3N2$ indexes. For models with 100  and 190 kK all observational data fall within the regions occupied by the models, while models with $T_{\rm eff}$=50\,kK do not reproduce well the  emission line ratios for the $N2$ index. This result agrees with that found by \citet{2021MNRAS.505.2087K} for UGC\,4805. Thus, the  photoionization model results
with $T_{\rm eff}$=50\,kK  for the $N2$ index
were not  taken into account in our study. The  values of O/H  and $U$
were derived using  linear interpolations between  the photoionization models following the procedure applied by  \citet{2021MNRAS.505.2087K}. In this way,  we derived three pairs of points for each object: ($N2$, O/H), ($O3N2$, O/H) and ($O3O2$, $\log U$). Then,  we obtained a set of point pairs for the sample and analysed the building of the calibrations.

\begin{figure*}
\centering
\begin{minipage}{.49\textwidth}
\includegraphics[width=0.9\textwidth]{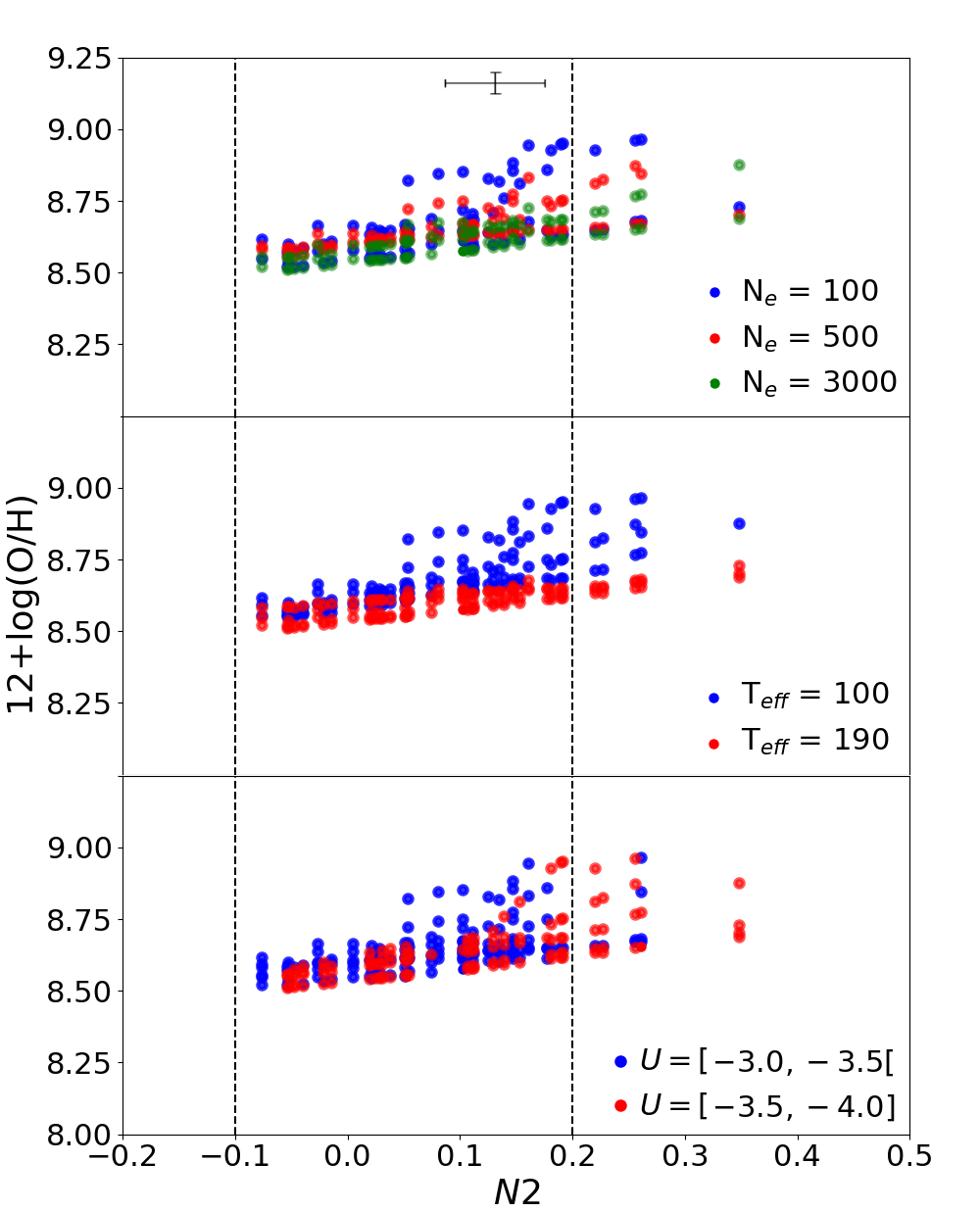}
\end{minipage}
\begin{minipage}{.49\textwidth}
\includegraphics[width=.9\textwidth]{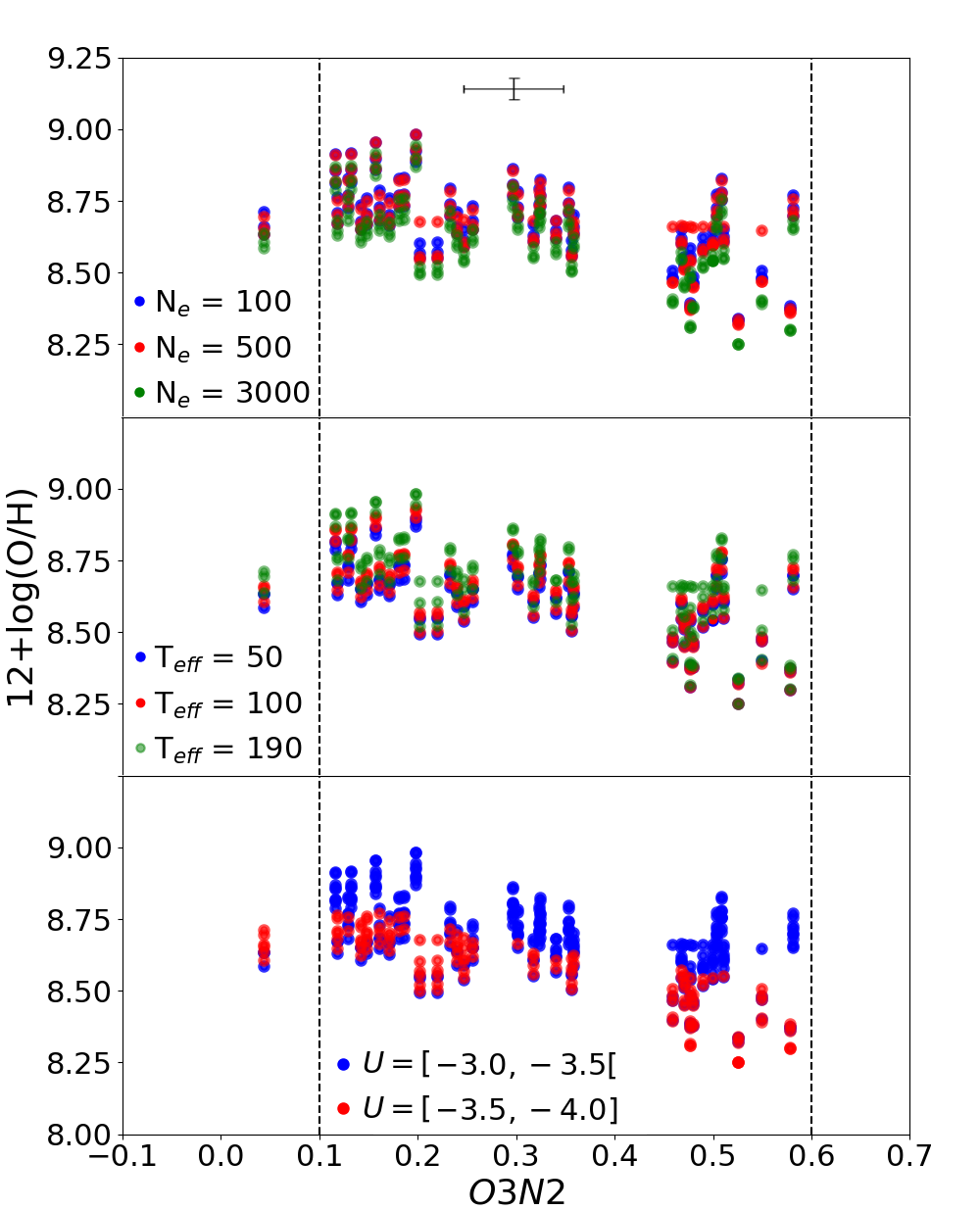}
\end{minipage}\caption{Left column: oxygen abundance versus the $N2$ index. Oxygen abundances were estimated interpolating the results from the models. Different colours are used to discriminate the estimated properties for the objects in our sample considering different $N_{\rm e}$ (upper panel),
$T_{\rm eff}$ (middle panel), and $\log (U)$ (bottom panel) values as indicated. Right column: same as the left column but for $O3N2$. Error bars in each panel represent the typical 0.1 dex error in the observational measurements of the
$N2$  and $O3N2$ indexes and the average error of 0.10 and 0.12 dex in the interpolated values, considering $N2$ and $O3N2$ respectively. The grey vertical lines correspond 
to the limits of the $-0.1<N2<0.2$ and $0.1<O3N2<0.6$ ranges,  which correspond to the bulk of the observational data.}
\label{comp}
\end{figure*}

That allowed us to analyse the dependency  of the  oxygen abundance with $N_{\rm e}$,
$T_{\rm eff}$, and $\log (U)$ as a function of the $N2$ and $O3N2$ indexes, which are shown
in Fig.~\ref{comp}. Data exhibited in this figure correspond to the interpolated model values obtained from Figs.\ \ref{grid_pagb_n2} and \ref{grid_pagb_o3n2} for each object, i.e., each galaxy has 6 points in each left panel of Fig.~\ref{comp} and 9 points in each right panel of Fig.~\ref{comp}.
We limited the analysis to the ranges of $-0.1<N2<0.2$ and $0.1<O3N2<0.6$,  which correspond to the bulk of the observational data.  For the $N2$ index (left panels of Fig.~\ref{comp}), the higher  oxygen abundance values were derived from lower values of $T_{\rm eff}$ and $N_{\rm e}$, without dependency on  $\log U$. On the other hand, for the $O3N2$ index,  the oxygen abundance  is dependent on the $N_{\rm e}$, $T_{\rm eff}$ and $\log U$, in that the higher oxygen abundance was derived from the lower values of density and higher values of  effective temperature and ionization parameter. However, as seen in Fig.~\ref{comp}, the average error in
the observational values of $N2$ and $O3N2$ produces uncertainties in the  abundance estimations  in the order of or even larger than those due to the variation of the nebular parameters (see also \citealt{2020MNRAS.492.5675C}).

\begin{figure*}
\centering
\begin{minipage}{.49\textwidth}
\includegraphics[width=0.9\textwidth]{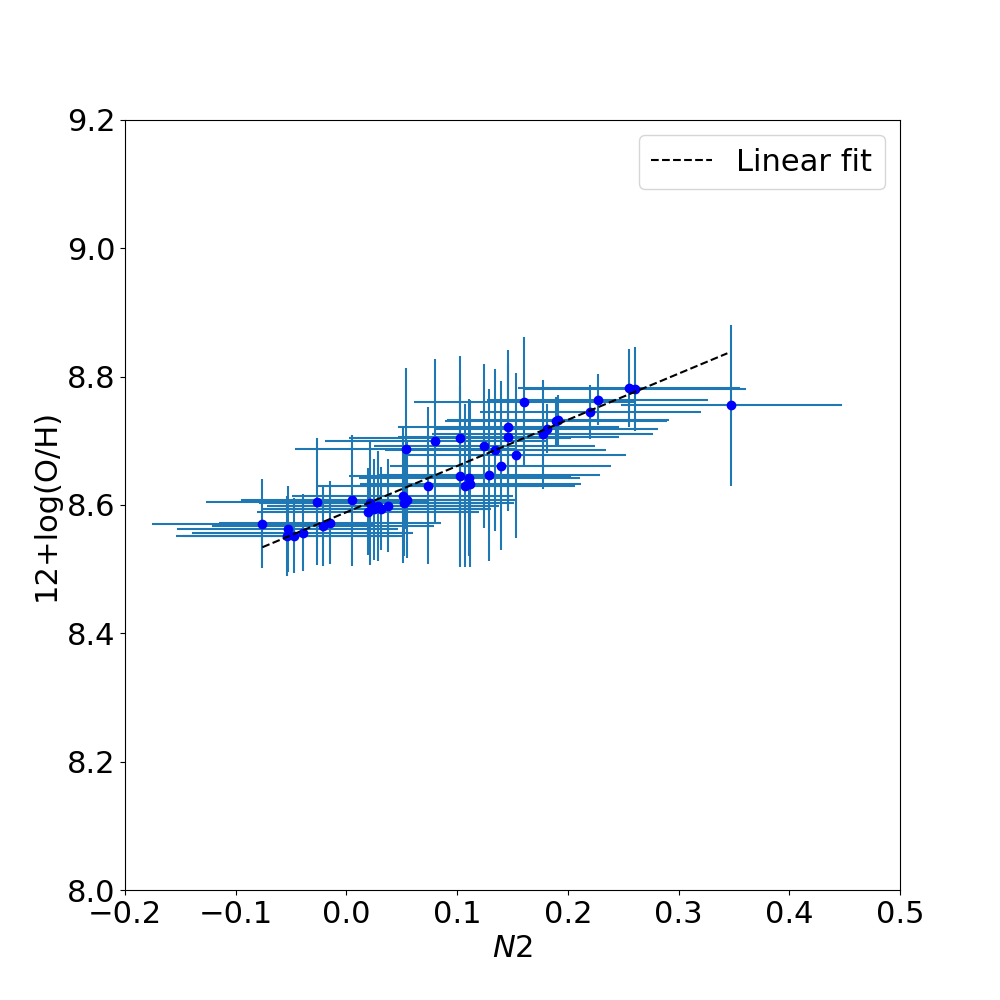}
\end{minipage}
\begin{minipage}{.49\textwidth}
\includegraphics[width=.9\textwidth]{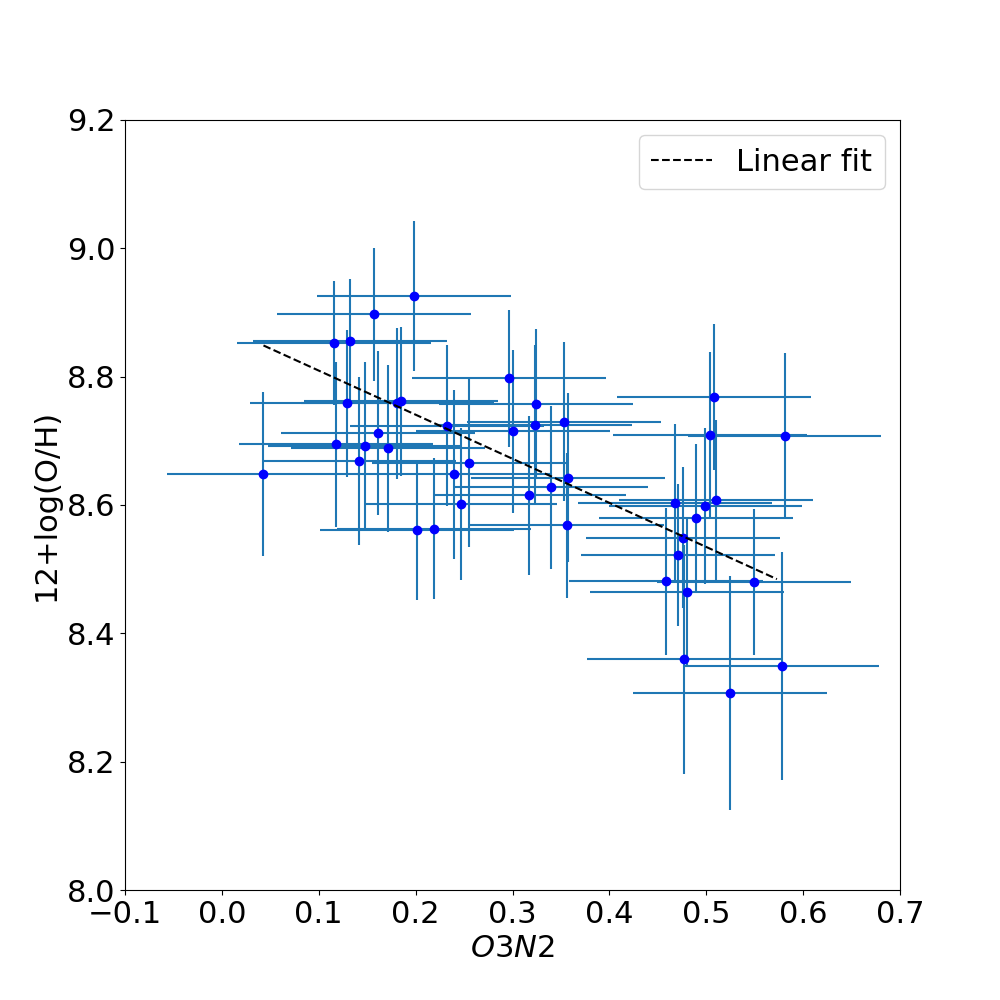}
\end{minipage}\caption{Left panel: oxygen abundances versus the $N2$ index. Points represent the average estimations from the  photoionization model results with errors. Right panel: Same as left panel but for oxygen abundance versus $O3N2$ index. Black line represents the linear fit given by eqs.\ \ref{eq_n2} and \ref{eq_o3n2} applying the ODR method, i.e., considering the errors, whose correlation coefficients are $R = 0.85$ and $R = 0.37$, respectively. }

\label{calib_N2_O3O2}
\end{figure*}

Therefore, using the orthogonal distance regression (ODR) method, which takes into account errors in both the x and y variables \citep{boggs},  we performed two unidimensional semi-empirical calibration considering all points, i.e., for each galaxy the assumed oxygen abundance value is the average of the interpolations for the six photoionization grid models from Fig.~\ref{grid_pagb_n2} and the nine photoionization grid models from Fig.~\ref{grid_pagb_o3n2}. 
Fig.~\ref{calib_N2_O3O2} presents the averaged  oxygen abundance values  versus the  $N2$ and $O3N2$ indexes (left and right panels, respectively). The
$N2$ index is well correlated with the oxygen abundance with low dispersion, while the $O3N2$ index also has  a linear correlation with the oxygen abundance, but with a higher dispersion. The derived linear calibrations are given by

\begin{equation}\label{eq_n2}
    12 + \log ({\rm O/H}) = 0.71 (\pm 0.03) N2 + 8.58 (\pm 0.01)
\end{equation}
\noindent and
\begin{equation}\label{eq_o3n2}
     12 + \log ({\rm O/H}) = -0.68 (\pm 0.11) O3N2 + 8.87 (\pm 0.03)
\end{equation}

\begin{figure}
\includegraphics*[width=0.45\textwidth]{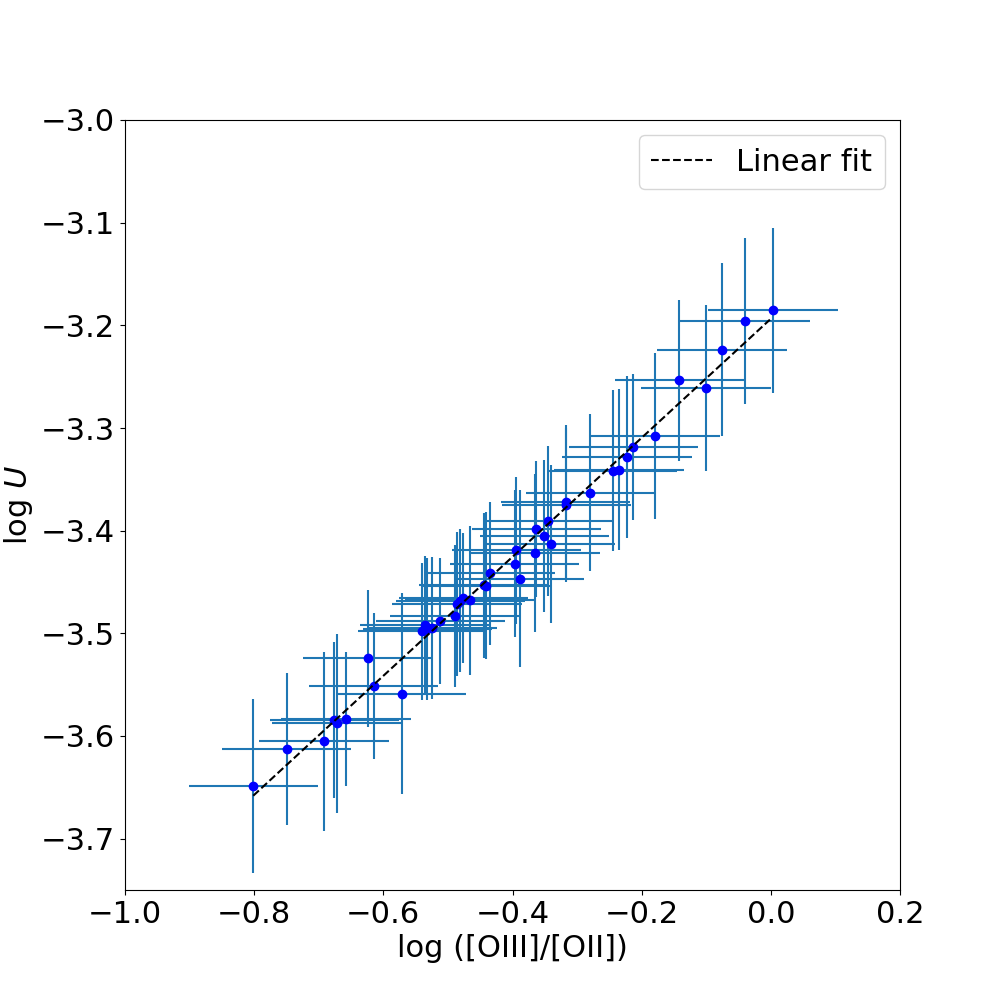}
\caption{Same ass Fig.~\ref{grid_pagb_n2}, but for $\log U$ versus
log([\ion{O}{iii}]$\lambda5007$/[\ion{O}{ii}]$\lambda3727$.
Line represents the linear fitting (Eq.~\ref{eqcalu}) whose correlation coefficient is $R=0.98$. }
\label{calu} 
\end{figure}

The interpolated values from Figs.~\ref{grid_pagb_n2}  and \ref{grid_pagb_o3n2}
can also be used to derive a calibration between the 
observed log([\ion{O}{iii}]$\lambda5007$/[\ion{O}{ii}]$\lambda3727$) and the ionization
parameter $U$, which is assumed as the average, for each object, of the interpolated model values. Fig~\ref{calu} contains these values, together with the obtained linear regression fit given by

\begin{equation}
\label{eqcalu}
    \log U= 0.57(\pm0.01) \:\rm  x \:-3.19(\pm0.01),
\end{equation}
where x=log([\ion{O}{iii}]$\lambda5007$/[\ion{O}{ii}]$\lambda3727$).
We did not find any
dependence of this relation on $N_{\rm e}$,  $T_{\rm eff}$  and O/H, similar to the results obtained by \citet{2020MNRAS.492.5675C} for Narrow Line Regions of Seyfert 2 galaxies.

%%%%%%%%%%%%%%%%%%%%%%%%%%%%%%%%%%%%5%%%%%%%%%%%%%%%
\section{Discussions}
\label{disc}

 In the chemical abundance determinations of the gas phase, the knowledge of the ionizing source  is fundamental, especially when  it is estimated using indirect methods. The nature of the ionizing source of LINERs is an open problem in 
astronomy, and  three mechanisms have been proposed as responsible for
the ionization: shocks \citep{1980A&A....87..152H}, accretion gas 
into a central black hole (AGN, \citealt{1983ApJ...269L..37H, 1983ApJ...264..105F, 1993ApJ...417...63H}), and hot stars (post-AGB stars, \citealt{10.1093/mnras/213.4.841, 1992ApJ...399L..27S, Taniguchi_2000}). 
  In our specific case, the sample is composed by objects with a LINER nucleus, and according to the WHAN diagram are classified as retired galaxies.
  %. Moreover, all galaxies of our sample present an extended ionization (up to $\sim 4$ kpc) from central regions. 
  We argued that the LINER ionization sources of these galaxies are probably post-AGB stars spread along the gas (see also \citealt{2021MNRAS.505.2087K}). Therefore, based  on this assumption, we proposed two semi-empirical calibrations
between the $N2$ and $O3N2$ line ratios and the metallicity, as well as a calibration   between  [\ion{O}{iii}]/[\ion{O}{ii}] ratio  and the ionization parameter $U$ derived from photoionization models assuming the ionizing sources are post-AGBs. 

\begin{figure*}
\centering
\begin{minipage}{.49\textwidth}
\includegraphics[width=0.9\textwidth]{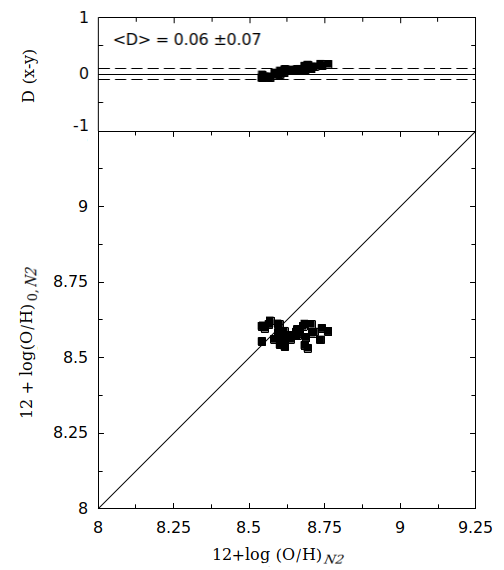}
\end{minipage}
\begin{minipage}{.49\textwidth}
\includegraphics[width=.9\textwidth]{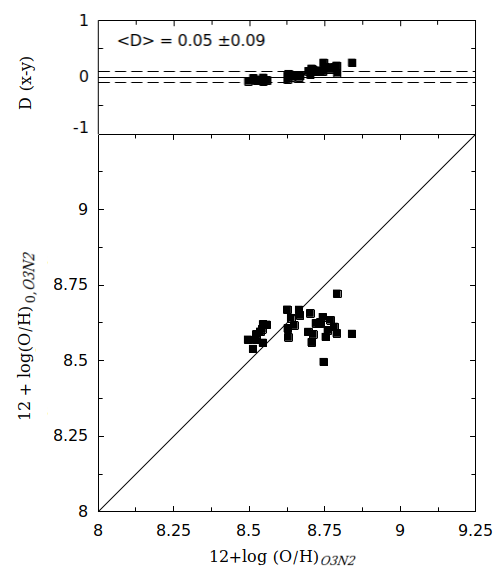}
\end{minipage}
\caption{Left bottom panel:  comparison between $\rm 12+log(O/H)$ obtained using the calibration ${\rm 12+log(O/H)}-N2$ proposed in this work and the one obtained through the extrapolation of the radial gradient (see Sec.~\ref{star_forming}). Left top panel: difference $<D>$ between the metallicity estimations based on our ${\rm 12+log(O/H)}-N2$ relation and the one obtained through the extrapolation using the $N2$ index. The average difference between these estimations is provided. Dashed lines indicate the uncertainty of $\pm 0.1$ dex assumed in $\rm 12+log(O/H)$  estimations via strong emission-line methods (\citealt{kewley01}).  Right panels: the same as the left panels but using the $O3N2$ index.}
\label{dif}
\end{figure*}

\begin{figure}
\includegraphics*[width=0.45\textwidth]{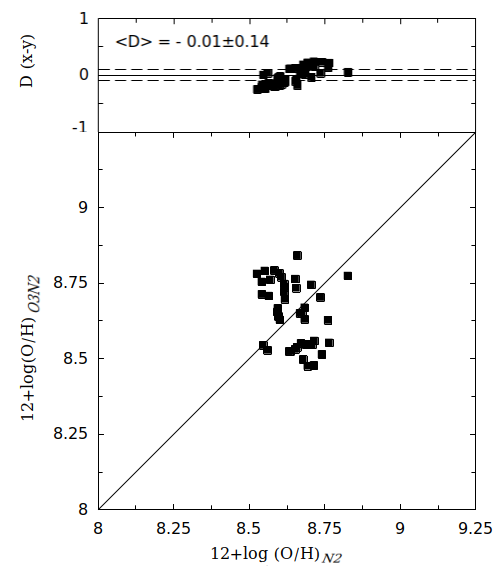}
\caption{Bottom panel: oxygen abundance estimations derived through our $O3N2$ calibration (Eq.~\ref{eq_o3n2}) plotted against the ones estimated through the $N2$ calibration (Eq.~\ref{eq_n2}). Solid line represents the equality between these oxygen estimations. Top panel: difference $<D>$ between both estimations. The average difference is indicated.} 
\label{comparation} 
\end{figure}

In Table~\ref{media}, we list for all sampled galaxies the oxygen abundance values obtained using  the calibrations proposed in this work  (see Sec. \ref{res})  and  those derived by  extrapolating the radial gradients  $\rm 12+log(O/H)_{0}$ (see Sec. \ref{star_forming}), as well as  the ionization parameter values obtained by using Eq.~\ref{eqcalu}.  Fig.~\ref{dif}
shows  $\rm 12+log(O/H)_{0}$ versus $\rm 12+log(O/H)$
through the $N2$ index (left panel) and those through the $O3N2$ index (right panel). Both  ${\rm 12+log(O/H)}-N2$ and ${\rm 12+log(O/H)}-O3N2$ relations produced higher oxygen abundance values than those derived by the oxygen abundance gradient extrapolation  method. 
%The average differences  $<D> = 0.06 \pm 0.08$ and $<D> = 0.06 \pm 0.07$, considering the $O3N2$ and $N2$  indexes, respectively. Nevertheless, these differences are at the same order of the accuracy of  0.18 and 0.16 dex of $O3N2$ and $N2$ calibrations of  \cite{marino}, respectively.
Taking into account the observational uncertainties ($\sim 0.1$ dex, \citealt{2003ApJ...591..801K}) and the accuracy of the theoretical and (semi)empirical calibrations ($\sim 0.1$ dex, e.g., \citealt{2002ApJS..142...35K}, \citealt{2020MNRAS.492.5675C}, \citealt{2021MNRAS.507..466D}), we can claim that the present estimated abundances through the different methods agree with each other. Therefore, these estimations support the validity of the semi-empirical calibration for LINER objects obtained in this work.
%Thus, they agreement within the uncertainties, giving an excellent support to the semi-empirical calibration for LINER objects obtained in this work. 
Using these calibrations,  we found that LINERs exhibit an oxygen abundance range $\rm 8.48 \: \la \: 12+log(O/H) \: \la  8.84$, with a mean value of $\rm 12+\log(O/H)=8.65$. Fig.~\ref{comparation} contains a comparison between the abundance values obtained using our two calibrations. 
$N2$ and $O3N2$ oxygen abundance estimations are in agreement. However, this relation has a high dispersion, and the  difference between both estimations (upper panel) exhibits a systematic linear behaviour. Therefore, considering these results and that the $O3N2$ calibration presents a higher dispersion than the $N2$ (see Fig. \ref{calib_N2_O3O2}), 
even if we are able to use both calibrations to estimate the central oxygen abundance of LINER galaxies, we recommend using the $N2$ calibration.

%We can see a very well agreement between $N2$ and $O3N2$ calibrations, then any of them could be used to estimated the metallicity abundance in LINERs. 

%In the left panel of Fig.~\ref{dif} the estimations for $N2$ index are plotted in bottom panel and the difference between them are plotted in the top panel. The average difference between both estimations is $D = 0.07 \pm 0.09$.  The average difference between the results obtained by our calibration and the extrapolation of gradients considering $O3N2$ index is $D = 0.15 \pm 0.12$, value that is slightly higher than the uncertainty of $\pm 0.1$ dex assumed by \cite{denicolo} in metallicity estimations via strong emission-line methods. 

\begin{table*}
\centering
\caption{Oxygen abundance values obtained through the $N2$ and $O3N2$ calibrations presented in Eqs.~\ref{eq_n2} and \ref{eq_o3n2}, and the mean values of the ionization parameter obtained through Eq.~\ref{eqcalu}. Extrapolated radial oxygen abundance values, ${\rm 12+log(O/H)}_{0}$, estimated from both $N2$ and $O3N2$ indexes (see Sec.~\ref{star_forming}) are also shown.}
\label{media}
\begin{tabular}{@{}lccccccc@{}}
\hline
Plate-IFU       &\multicolumn{2}{c}{12+log(O/H)}  &     $<\log U>$                           &	     & \multicolumn{2}{c}{12+log(O/H)$_{0}$}    \\	  	      
\cline{2-3}
\cline{6-7}
%\cline{9-10}
\noalign{\smallskip}
&    $N2$       &  $O3N2$         	  &	     &       &    $N2$      &	  $O3N2$     &         	    \\  		
\noalign{\smallskip}				  	  			  	    							  			      
7495-12704    &     8.62 $\pm$0.02&	8.74 $\pm$ 0.05   &	 $-$3.19$\pm$ 0.01	    &	 &  8.54 $\pm$   0.02&  8.63$\pm$  0.02 \\
7977-3704     &     8.57 $\pm$0.01&	8.71 $\pm$ 0.06   &	 $-$3.47$\pm$ 0.02	    &	 &  8.61 $\pm$   0.02&  8.56$\pm$  0.03 \\
7977-12703    &     8.54 $\pm$0.00&	8.75 $\pm$ 0.05   &	 $-$3.46$\pm$ 0.01	    &	 &  8.60 $\pm$   0.01&  8.58$\pm$  0.02 \\
7990-6103     &     8.55 $\pm$0.01&	8.54 $\pm$ 0.08   &	 $-$3.55$\pm$ 0.02	    &	 &  8.61 $\pm$   0.02&  8.60$\pm$  0.03 \\
7990-12704    &     8.59 $\pm$0.01&	8.67 $\pm$ 0.06   &	 $-$3.37$\pm$ 0.01	    &	 &  8.61 $\pm$   0.03&  8.67$\pm$  0.05 \\
8083-12704    &     8.58 $\pm$0.01&	8.79 $\pm$ 0.04   &	 $-$3.32$\pm$ 0.01	    &	 &  8.56 $\pm$   0.02&  8.72$\pm$  0.03 \\
8131-9102     &     8.71 $\pm$0.06&	8.55 $\pm$ 0.08   &	 $-$3.62$\pm$ 0.02	    &	 &  8.58 $\pm$   0.04&  8.56$\pm$  0.09 \\
8140-12703    &     8.62 $\pm$0.02&	8.72 $\pm$ 0.05   &	 $-$3.57$\pm$ 0.02	    &	 &  8.56 $\pm$   0.02&  8.62$\pm$  0.02 \\
8243-9102     &     8.83 $\pm$0.12&	8.77 $\pm$ 0.05   &	 $-$3.50$\pm$ 0.02	    &	 &  -		     &  -	     \\
8243-12701    &     8.69 $\pm$0.04&	8.55 $\pm$ 0.08   &	 $-$3.46$\pm$ 0.01	    &	 &  8.57 $\pm$   0.02&  8.62$\pm$  0.02 \\
8247-3701     &     8.60 $\pm$0.02&	8.78 $\pm$ 0.04   &	 $-$3.42$\pm$ 0.01	    &	 &  8.61 $\pm$   0.02&  8.61$\pm$  0.05 \\
8249-12704    &     8.64 $\pm$0.03&	8.52 $\pm$ 0.09   &	 $-$3.21$\pm$ 0.01	    &	 &  8.56 $\pm$   0.02&  8.59$\pm$  0.03 \\
8252-12702    &     8.71 $\pm$0.02&	8.48 $\pm$ 0.09   &	 $-$3.58$\pm$ 0.01	    &	 &  -	       &  -		     \\
8254-3704     &     8.62 $\pm$0.01&	8.70 $\pm$ 0.06   &	 $-$3.44$\pm$ 0.01	    &	 &  8.54 $\pm$   0.02&  8.59$\pm$  0.02 \\
8257-1902     &     8.57 $\pm$0.01&	8.76 $\pm$ 0.05   &	 $-$3.44$\pm$ 0.02	    &	 &  8.62 $\pm$   0.03&  8.60$\pm$  0.04 \\
8258-12704    &     8.74 $\pm$0.05&	8.51 $\pm$ 0.09   &	 $-$3.65$\pm$ 0.01	    &	 &  8.60 $\pm$   0.04&  8.54$\pm$  0.06 \\
8259-9102     &     8.60 $\pm$0.04&	8.63 $\pm$ 0.07   &	 $-$3.50$\pm$ 0.01	    &	 &  8.54 $\pm$   0.02&  8.61$\pm$  0.02 \\
8313-9102     &     8.71 $\pm$0.04&	8.74 $\pm$ 0.05   &	 $-$3.38$\pm$ 0.01	    &	 &  8.61 $\pm$   0.01&  8.64$\pm$  0.03 \\
8313-12705    &     8.68 $\pm$0.04&	8.63 $\pm$ 0.07   &	 $-$3.33$\pm$ 0.01	    &	 &  8.61 $\pm$   0.02&  8.58$\pm$  0.05 \\
8318-12703    &     8.68 $\pm$0.06&	8.67 $\pm$ 0.06   &	 $-$3.29$\pm$ 0.02	    &	 &  8.54 $\pm$   0.02&  8.65$\pm$  0.02 \\
8320-9102     &     8.67 $\pm$0.03&	8.65 $\pm$ 0.07   &	 $-$3.32$\pm$ 0.01	    &	 &  8.58$\pm$	 0.02&  8.62$\pm$  0.03 \\
8332-12705    &     8.72 $\pm$0.02&	8.56 $\pm$ 0.08   &	 $-$3.54$\pm$ 0.01	    &	 &  8.58 $\pm$   0.02&  8.62$\pm$  0.03 \\
8330-9102     &     8.65 $\pm$0.02&	8.76 $\pm$ 0.05   &	 $-$3.25$\pm$ 0.01	    &	 &  -		     &  -	    \\
8332-6103     &     8.63 $\pm$0.01&	8.52 $\pm$ 0.09   &	 $-$3.37$\pm$ 0.02	    &	 &  8.57$\pm$	 0.02&  8.57$\pm$	0.04 \\
8440-12704    &     8.62 $\pm$0.08&	8.75 $\pm$ 0.05   &	 $-$3.39$\pm$ 0.01	    &	 &  8.59 $\pm$   0.01&  8.49$\pm$  0.06 \\
8481-1902     &     8.59 $\pm$0.04&	8.65 $\pm$ 0.06   &	 $-$3.46$\pm$ 0.02	    &	 &  -		&  -	     \\
8482-12703    &     8.76 $\pm$0.02&	8.63 $\pm$ 0.07   &	 $-$3.41$\pm$ 0.02	    &	 &  8.59 $\pm$   0.02&  8.67$\pm$  0.03 \\
8549-3703     &     8.68 $\pm$0.07&	8.50 $\pm$ 0.09   &	 $-$3.49$\pm$ 0.02	    &	 &  8.60 $\pm$   0.01&  8.57$\pm$  0.02 \\
8550-6103     &     8.61 $\pm$0.03&	8.77 $\pm$ 0.05   &	 $-$3.47$\pm$ 0.01	    &	 &  8.56 $\pm$   0.01&  8.63$\pm$  0.02 \\
8550-12704    &     8.74 $\pm$0.03&	8.70 $\pm$ 0.06   &	 $-$3.52$\pm$ 0.02	    &	 &  8.56 $\pm$   0.02&  8.66$\pm$  0.04 \\
8550-12705    &     8.66 $\pm$0.00&	8.54 $\pm$ 0.08   &	 $-$3.41$\pm$ 0.01	    &	 &  8.57 $\pm$   0.03&  8.60$\pm$  0.05 \\
8552-9101     &     8.66 $\pm$0.01&	8.84 $\pm$ 0.03   &	 $-$3.56$\pm$ 0.02	    &	 &  8.59 $\pm$   0.02&  8.59$\pm$  0.02 \\
8601-12705    &     8.54 $\pm$0.04&	8.71 $\pm$ 0.06   &	 $-$3.40$\pm$ 0.01	    &	 &  8.55$\pm$	 0.02&  8.59$\pm$	0.03 \\
8588-9101     &     8.55 $\pm$0.01&	8.79 $\pm$ 0.04   &	 $-$3.48$\pm$ 0.01	    &	 &  8.60 $\pm$   0.01&  8.59$\pm$  0.02 \\
8138-3702     &     8.67 $\pm$0.03&	8.55 $\pm$ 0.08   &	 $-$3.49$\pm$ 0.02	    &	 &  -		&  -	     \\
8138-9101     &     8.60 $\pm$0.01&	8.64 $\pm$ 0.07   &	 $-$3.44$\pm$ 0.01	    &	 &  8.58$\pm$	 0.02&  8.64$\pm$	0.02	  \\
8482-3704     &     8.66 $\pm$0.00&	8.73 $\pm$ 0.05   &	 $-$3.58$\pm$ 0.01	    &	 &  8.58$\pm$	 0.02&  8.62$\pm$	0.03	  \\
8482-9101     &     8.69 $\pm$0.08&	8.48 $\pm$ 0.09   &	 $-$3.23$\pm$ 0.01	    &	 &  8.53$\pm$	 0.06&  -	      \\
8554-1902     &     8.56 $\pm$0.04&	8.53 $\pm$ 0.09   &	 $-$3.27$\pm$ 0.01	    &	 &  -		     &  -	    \\
8603-12703    &     8.53 $\pm$0.03&	8.78 $\pm$ 0.04   &	 $-$3.31$\pm$ 0.01	    &	 &  -		     &  -	    \\
8604-12703    &     8.76 $\pm$0.04&	8.55 $\pm$ 0.08   &	 $-$3.40$\pm$0.01	    &	 &	 -		  &  -  	 \\
8604-6102     &     8.68 $\pm$0.03&	8.65 $\pm$ 0.07   &	 $-$3.35$\pm$0.02	    &	 &	 -		  &  -  	 \\
8606-3702     &     8.65 $\pm$0.03&	8.53 $\pm$ 0.08   &	 $-$3.39$\pm$0.01	    &	 &		 -		  &  -  	 \\

\hline						
\end{tabular}					
\end{table*}

\begin{figure}
\includegraphics*[width=0.45\textwidth]{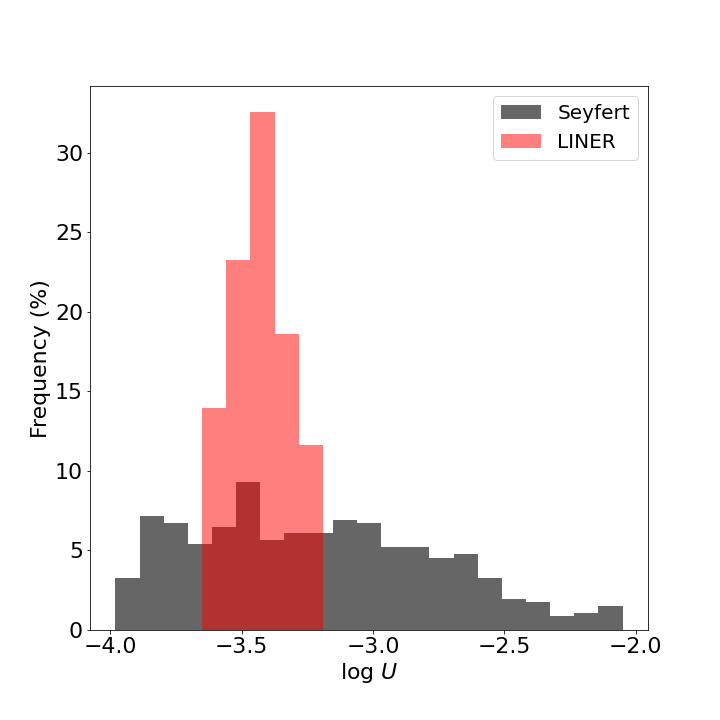}
\caption{Histogram containing the logarithm of the ionization parameters of our objects derived through Eq.~\ref{eqcalu} and the values derived for a sample of  Seyfert 2 galaxies studied by \citet{2020MNRAS.492.5675C}.}  
\label{log} 
\end{figure}

For the ionization parameter, it is well known that LINERs have a lower ionization parameter than Seyferts (see \citealt{1983ApJ...264..105F,2006MNRAS.372..961K}).  Fig.~\ref{log} compare the estimated values of the logarithm of the ionization parameter ($\log U$) of our sample of LINERs with those derived for a sample of Seyfert 2 galaxies studied by \cite{2020MNRAS.492.5675C}. The values of 
$\log U$  are in the range from $-4$ to $-2.0$ and from $-3.6$ to $-3.2$, for the  Seyfert 2 and LINERs, respectively. The mean value of 
$\log U$ is about  $-3.21$ ($\sigma = 0.45$) and  $-3.42$ ($\sigma = 0.11$), for the Seyfert 2 and LINER galaxies, respectively. This difference represents a factor of almost 2 on a linear scale. Hence, although some Seyferts 2 have lower ionization parameters, they extend to much higher values than the ones estimated for the LINERs. 
This large variation in $\log U$ was also reported by \cite{1984A&A...135..341S} for Seyfert nuclei. This author presented a series of diagrams in which different line intensity ratios are plotted for a sample of Seyfert 2 galaxies and, found that objects with low density clouds ( $ \rm < 10^4 cm^3$) present $\log U$ that varies from  $-4$ to $-2$.

 %These authors found for their sample of Seyfert 2 that  $\log U$ is in the range  of while for our sample of LINERs it is in the range {\bf with average of $-3.42$ and $\sigma = 0.11$)}.  Although  some Seyferts have lower ionization parameters (about the same range of the LINERs), they extend to much higher values than ones estimated for the LINERs as observed in Fig.~\ref{log}. \textbf{A similar result was found by \cite{2021MNRAS.505.2087K}. These authors compared a sample of 38 LINER galaxies, obtained by \cite{1993ApJ...417...63H}, \cite{2001ApJ...554..240E},  \cite{2010A&A...519A..40A} and \cite{2018ApJ...864...90M} with a sample of Seyfert galaxies obtained by \cite{2020MNRAS.492.5675C}. They found that these LINER objects tend to have lower $U$ values than Seyfert galaxies, although some Seyfert galaxies have similar $U$ values to those derived for the LINER sample (see.}

\section{Conclusion}
\label{conc}

Using optical data of 43 LINER galaxies obtained from the MaNGA survey, we proposed, for the first time, two semi-empirical calibrations based on photoionization models to estimate the oxygen abundance of this class of objects, as a  function of the $N2$ and $O3N2$ emission-line intensity ratios. Due to the nuclei of the objects in our sample classified as RGs according to the WHAN diagnostic diagram,  we argue that these LINERs are probably ionized by post-AGB stars. Therefore, to derive the calibrations, we built photoionization models using the  {\sc cloudy} code considering post-AGB stars with three different effective temperatures (50, 100, 190 kK)  as the ionizing sources. %We find that models involving post-AGB stars with effective temperature of 50 kK do not reproduce the observed $N2$ emission-line ratios  for any of the studied objects. However, for $O3N2$ index we not found this situation, i.é., models with temperature of 50 kK could well reproduce the observed emission-line ratio. \sout{with different effective temperatures.}  
Using the calibrations proposed in this work, we found that LINERs exhibit an oxygen abundance range $\rm 8.48 \: \la \: 12+log(O/H) \: \la  8.84$, with a mean value of $\rm 12+\log(O/H)=8.65$. We compared the results produced by both calibrations and found they are in agreement. Comparing the results produced by the calibrations, taking into account the observational and theoretical errors, we found good agreement. Considering this result and that the  ${\rm 12+log(O/H)}-N2$ calibration presents a much smaller dispersion than the ${\rm 12+log(O/H)}-O3N2$ calibration, we recommend the use of the $N2$ index to estimate the oxygen abundances of LINERs. We compared the metallicities values produced by the proposed calibrations, with those derived by extrapolating the disk oxygen abundance gradients to the centre of the galaxies, finding that they are in good agreement. We also derived a calibration between the logarithm of the ionization parameter and the $[\ion{O}{III}]/[\ion{O}{II}]$ emission-line ratio.

%. We also derived a calibration between the logarithm of ionization parameter and the line ratio $[\ion{O}{III}]/[\ion{O}{II}]$. %{\bf We compared the computed values of the ionization parameter of our galaxies  with  a sample of Seyfert galaxies.} As expected, we found that the  ionization parameter of LINERs  are systematically lower than Seyferts galaxies. 

\section*{Acknowledgements}
CBO is grateful to the Fundação de Amparo à Pesquisa do Estado de São Paulo (FAPESP) for the support under grant 2019/11934-0 and to the Coordenação de Aperfeiçoamento de Pessoal de Nível Superior (CAPES).  ACK thanks  FAPESP for the support grant 	2020/16416-5 and the Conselho Nacional de Desenvolvimento Científico e Tecnológico (CNPq).  JAHJ acknowledges support from FAPESP, process number 2021/08920-8. OLD is grateful to FAPESP and CNPq. IAZ acknowledges support by the National Academy of Sciences of Ukraine under the Research Laboratory Grant for young scientists No. 0120U100148. AFM gratefully acknowledges support from CAPES.

{\it Software:} {\sc astroplotlib} \citep{astroplotlib, hernandez13, hernandez15}, {\sc astropy} \citep{astropyI, astropyII},
{\sc scipy} \citep{scipy}, {\sc numpy} \citep{numpy} and {\sc matplotlib} \citep{matplotlib}.

\section{Data Availability}

The data underlying this article will be shared on reasonable request
to the corresponding author.

%%%%%%%%%%%%%%%%%%%%%%%%%%%%%%%%%%%%%%%%%%%%%%%%%%

%%%%%%%%%%%%%%%%%%%% REFERENCES %%%%%%%%%%%%%%%%%%

% The best way to enter references is to use BibTeX:

\bibliographystyle{mnras}
\bibliography{example} % if your bibtex file is called example.bib

% Alternatively you could enter them by hand, like this:
% This method is tedious and prone to error if you have lots of references
%\begin{thebibliography}{99}
%\bibitem[\protect\citeauthoryear{Author}{2012}]{Author2012}
%Author A.~N., 2013, Journal of Improbable Astronomy, 1, 1
%\bibitem[\protect\citeauthoryear{Others}{2013}]{Others2013}
%Others S., 2012, Journal of Interesting Stuff, 17, 198
%\end{thebibliography}

%%%%%%%%%%%%%%%%%%%%%%%%%%%%%%%%%%%%%%%%%%%%%%%%%%

% Don't change these lines
\bsp	% typesetting comment
\label{lastpage}
\end{document}